\documentclass[twocolumn]{aastex631}

\usepackage{lineno}
\usepackage{txfonts} 
\usepackage{ulem}
\usepackage{xcolor}
\usepackage{hyperref}

\newcommand{\teff}{\mbox{T$_{\rm eff}$}}
\newcommand{\logg}{\mbox{log~{\it g}}}

\newcommand{\x}{\mbox{$\Delta_{\tiny{\mathrm{F275W,F814W}}}$}}
\newcommand{\y}{\mbox{$\Delta_{\tiny{C~\mathrm{ F275W,F336W,F438W}}}$}}

\accepted{February, 13 2024}

\shorttitle{A $JWST$ project overview on 47\,Tucanae} 
\shortauthors{A.\,F. Marino, et al.}

\begin{document}

\title{A $JWST$ project on 47~Tucanae.
Overview, photometry and early spectroscopic results of M dwarfs, and observation of brown dwarfs 
\footnote{
The main dataset presented in this article was obtained from the
Mikulski Archive for Space Telescopes (MAST) at the Space Telescope
Science Institute. The specific observations analyzed can be accessed
via \doi{10.17909/6t82-4360}.
}
}

\author{A.\ F.\,Marino} 
\affiliation{Istituto Nazionale di Astrofisica- Osservatorio Astronomico di Padova, Vicolo dell'Osservatorio 5, Padova, IT-35122} 
\affiliation{Istituto Nazionale di Astrofisica - Osservatorio Astrofisico di Arcetri, Largo Enrico Fermi, 5, Firenze, IT-50125}
\author{A.\ P.\,Milone}
\affiliation{Dipartimento di Fisica e Astronomia ``Galileo Galilei'' - Univ. di Padova, Vicolo dell'Osservatorio 3, Padova, IT-35122}
\affiliation{Istituto Nazionale di Astrofisica - Osservatorio Astronomico di Padova, Vicolo dell'Osservatorio 5, Padova, IT-35122} 
\author{M.\ V.\,Legnardi}
\affiliation{Dipartimento di Fisica e Astronomia ``Galileo Galilei'' - Univ. di Padova, Vicolo dell'Osservatorio 3, Padova, IT-35122}
\author{A.\,Renzini}
\affiliation{Istituto Nazionale di Astrofisica - Osservatorio Astronomico di Padova, Vicolo dell'Osservatorio 5, Padova, IT-35122} 
\author{E.\,Dondoglio}
\affiliation{Istituto Nazionale di Astrofisica - Osservatorio Astronomico di Padova, Vicolo dell'Osservatorio 5, Padova, IT-35122} 
\author{Y.\,Cavecchi}
\affiliation{Departament de Fisica, EEBE, Universitat Politecnica de Catalunya, Av. Eduard Maristany 16, E-08019 Barcelona, Spain} 
\author{G.\,Cordoni}
\affiliation{Research School of Astronomy and Astrophysics, Australian National University, Canberra, ACT 2611, Australia}
\author{A.\,Dotter}
\affiliation{Department of Physics and Astronomy, Dartmouth College, 6127 Wilder Laboratory, Hanover, NH 03755, USA}
\author{E.P.\,Lagioia}
\affiliation{South-Western Institute for Astronomy Research, Yunnan University, Kunming, 650500 P.R.China}
\author{T.\,Ziliotto}
\affiliation{Dipartimento di Fisica e Astronomia ``Galileo Galilei'' - Univ. di Padova, Vicolo dell'Osservatorio 3, Padova, IT-35122}
\author{M.\,Bernizzoni}
\affiliation{Dipartimento di Fisica e Astronomia ``Galileo Galilei'' - Univ. di Padova, Vicolo dell'Osservatorio 3, Padova, IT-35122}
\author{E.\,Bortolan}
\affiliation{Dipartimento di Fisica e Astronomia ``Galileo Galilei'' - Univ. di Padova, Vicolo dell'Osservatorio 3, Padova, IT-35122}
\author{M.\,G.\,Carlos}
\affiliation{Theoretical Astrophysics, Department of Physics and Astronomy, Uppsala University, Box 516, SE-751 20 Uppsala, Sweden}
\author{S.\,Jang}
\affiliation{Center for Galaxy Evolution Research and Department of Astronomy, Yonsei University, Seoul 03722, Korea}
\author{A.\,Mohandasan}
\affiliation{Dipartimento di Fisica e Astronomia ``Galileo Galilei'' - Univ. di Padova, Vicolo dell'Osservatorio 3, Padova, IT-35122}
\author{F.\,Muratore}
\affiliation{Dipartimento di Fisica e Astronomia ``Galileo Galilei'' - Univ. di Padova, Vicolo dell'Osservatorio 3, Padova, IT-35122}
\author{M.\,Tailo}
\affiliation{Dipartimento di Fisica e Astronomia Augusto Righi, Università degli Studi di Bologna, Via Gobetti 93/2, 40129, Bologna, Italy}

\correspondingauthor{A.\ F.\,Marino}
\email{anna.marino@inaf.it}

\begin{abstract}
The James Webb Space Telescope ($JWST$) observations have been
demonstrated to be efficient in detecting globular clusters' (GCs)
multiple stellar populations in the low-mass regime of M dwarfs. We
present an overview, and first results, of different projects that can
be explored by using the $JWST$ observations gathered under the GO2560
for 47\,Tucanae, a first program entirely devoted to the investigation
of multiple populations in very low mass stars, which includes
spectroscopic data for the faintest GC stars for which spectra are
available. Our color-magnitude diagram (CMD) shows some substructures
for ultracool stars, including gaps and breaks in slope. In
  particular, we observe both a gap and a minimum in the F322W2
  luminosity function less than one magnitude apart, and discuss
  which one could be associated with the H-burning limit. We detect
stars fainter than this minimum, very likely the brown dwarfs. We
corroborate the ubiquity of the multiple populations across different
masses, from $\sim 0.1~{\rm M_{\odot}}$ up to red giants ($\sim
0.8~{\rm M_{\odot}}$). The oxygen range inferred from the M dwarfs,
both from the CMD and from the spectra of two M dwarfs
associated with different populations, is similar to that observed in
giants.
We have not detected any difference between the fractions of stars in distinct populations
across stellar masses $\gtrsim 0.1~{\rm M_{\odot}}$.
This work demonstrates the $JWST$'s capability in uncovering multiple populations within M
dwarfs and illustrates the possibility to analyse very low-mass stars
in GCs approaching the H-burning limit and the brown-dwarf
sequence. 
\end{abstract}


\section{Introduction}\label{sec:intro}

The majority of Globular Clusters (GCs) harbors diverse stellar
populations characterized by distinct chemical compositions. This
includes a first population (1P) of stars with compositions akin to
field stars exhibiting comparable metallicity, alongside one or more
'second' populations (2Ps) enriched in helium, nitrogen, aluminum, and
sodium, while being depleted in carbon and oxygen \citep[e.g.\,][]{kraft1994a,
  bastian2018a, gratton2019a, milone2022a}. 

Since first discoveries, the intricate complexities underlying the
coexistence of distinct stellar populations in GCs have posed a
challenge to our understanding of both stellar evolution and star 
formation in the early Universe. 
In the past decades, a huge observational effort has been devoted to
the study of multiple populations in GCs and several physical
processes have been 
proposed for their formation. However, a consensus regarding the
origin of the phenomenon is still missing  
\citep[see reviews by][and references therein]{renzini2015a,
  bastian2018a, milone2022a}.  

The current understanding of the "observed framework" provides two
main competing scenarios. According to the multiple generations
scenarios, multiple populations are attributed to distinct bursts of
star formation at different epochs. 
Second-generation (2P) stars are born from material processed and
ejected by the first-generation \citep[1P, e.g.\,][]{dantona1983a,
  cottrell1981a, renzini2022a}. Various kinds of 
1P polluters have been proposed, including massive binary, Asymptotic
Giant Branch (AGB), rotating, and supermassive stars
\citep[e.g.][]{ventura2001a, decressin2007a, demink2009a, krause2013a,
  denissenkov2014a, dantona2016a, calura2019a, lacchin24}.  
However, regardless of proposed polluters, the amount of available
material for 2P star formation is only a small fraction of the total
present-day mass of 1P stars, a challenge  known as the "mass
budget problem". 
To cope with  it, the 1P is required to be
substantially more massive at its formation and the progenitors
  of GCs should have lost a 
large fraction of it, thus providing a significant contribution to the
mass of the Galactic Halo. In this context, it has been argued
  that GCs would have formed inside dwarf galaxies as a result of a
  cooling catastrophe, with a fraction of the dwarf itself having
  contributed material for the formation of 2Ps
  \citep{renzini2022a}. As opposed to a compact, massive progenitor
  that would not lose much mass, a dense GC embedded in
  an extended envelope would easily survive tidal interaction while loosing
most of the envelope itself, e.g., as suggested by N-body simulations
\citep{lacchin24}.

Alternatively, it has been suggested that a fraction of stars from the same
generation successively accreted material processed and ejected by
massive stars of their own generation \citep{bastian2013a, gieles2018a}.
This scenario was proposed as an attempt to cope with the
mass budget issue, assuming that a much lesser amount of
  processed material would be required to produce the 2Ps, compared to
  the multiple stellar generations case. 
However, if only a small amount of accreted material was sufficient to produce the 2Ps,
then one would expect this material to be mixed over the whole
envelope as stars reach the red giant branch (RGB), thus producing a
difference in composition with respect to main sequence (MS) stars,
which is not observed. Even more fundamentally, as 2P stars are enriched in
helium, the putative accreted material would have a higher mean
molecular weight compared to underlying layers, thus causing 
Rayleigh\,Taylor instability to mix the accreted material with most of
the star. Thus, accretion seems not to solve the mass budget problem, as it 
hardly account for the {\it discreteness} of multiple populations \citep{renzini2015a}.
More recently, the occurrence of stellar mergers within forming
binary-rich globular clusters has been suggested as an explanation for
multiple populations observed in GCs \citep{wang2020a}. 

The two scenarios (multiple stellar generations and accretion)
result in different expectations for the population pattern across
stellar masses. 
Based on the multiple generations scenario,
the population pattern should be identical for
stars with different masses, which means that both high-mass and
low-mass stars formed during each burst of star formation should
exhibit similar chemical compositions. 
On the other hand, in the accretion scenario the amount of accreted material depends on
stellar mass, and it is proportional to the square of the stellar mass
($e.g., \propto \mathcal{M}^{2}$) \citep[in the case of
Bondi-like accretion][]{bondi1944a}. Being very-low 
mass stars less efficient in accreting polluted material, the
differences in helium, carbon, nitrogen, and oxygen (O) abundances
commonly observed among red giants, would 
systematically decrease in the M dwarf domain. 

The predictions of the two scenarios also diverge concerning the mass
function (MF) of the Multiple Populations (MPs). 
If the 2P grew out of Bondi accretion, the MF of 2P should be
significantly flatter compared to that of 1P \citep{Ballesteros-Paredes2015}.
In contrast, if there were multiple star-formation episodes, a much
less pronounced difference, if any, is expected between the MF slopes of 1P and 2P stars.
Being the multiple stellar population pattern across different masses
a key ingredient to constrain the formation scenarios, the comparison
of stellar population properties in the very-low mass regime of M
dwarfs with those, already known, of red giants, holds the 
promise of shedding light on the origin of this enigmatic phenomenon.

First analyses of multiple stellar populations below the MS knee have been conducted with data from the Hubble Space Telescope ($HST$) for the GC NGC~2808, which already showed the endurance of different stellar populations among low mass stars \citep{milone2012c}. Other GCs were successively investigated including NGC\,6121 \citep[M~4][]{milone2014}, Omega~Centauri \citep{milone2017b}, NGC\,6752 \citep{milone2019}.
This work highlights the presence of similar chemical variations among low and higher mass stars supporting the multiple generations scenario.
\citet{dondoglio2022a} presents an analysis of a sample of GCs below the MS knee, and provided for the first time, their MFs of multiple populations in NGC\,2808 and M~4 along a wide range of stellar masses, from $\sim$0.2 to $\sim$0.8~$\rm{M_{\odot}}$, finding that the fraction of MPs does not depend on the stellar mass, thus further challenging the accretion scenario.
First evidence of multiple stellar populations in very low mass stars from $JWST$ data have been reported fo  47~Tucanae \citep{milone2023a}, M\,92 \citep{ziliotto2023a}, and NGC\,6440 \citep{cadelano2023}.
In this paper we exploit deep $JWST$ data to reach well below the hydrogen-buring limit in 47~Tucanae, and into the brown dwarf regime.

This work provides an overview of a project, mainly based on
observations gathered under the $JWST$ program GO-2560. The project aims at the analysis of multiple stellar populations among very low mass stars, in the domain of M dwarfs, in the GC NGC~104 (47~Tucanae).
Since the seventies, the presence of multiple stellar populations with different light-element abundances in this GC has been strongly established from the analysis of bright RGB stars \citep[e.g.][]{dickens1979a, bell1983a, brown1990a, briley1991a}.
Nowadays, we know that 47\,Tucanae hosts two main stellar populations of 1P and 2P stars, which have been photometrically detected along the main evolutionary phases, including the MS, SGB, RGB, HB, and AGB \citep[e.g.][]{anderson2009a, milone2012a, lagioia2021a, jang2022a, tailo2020a, lee2022a}, and correspond to stars with different light-element abundances \citep{marino2019a}. 
A range of $\sim$0.4~dex in [C/Fe], $\sim$0.5~dex in [O/Fe], and $\sim$1.0~dex in [N/Fe], \citep{carretta2009a, marino2016a, dobrovolskas2014a} has been observed, whereas helium spans an interval of $\delta Y=$0.05 in mass fraction \citep{milone2018a}. 
Astrometric plus spectroscopic analysis, mostly based on RGB stars or bright MS stars, 
shows that these stellar populations exhibit different radial distributions and internal kinematics \citep[e.g.][]{richer2013a, milone2018b, cordoni2020a}.

In this work, we exploit the $JWST$ capabilities to explore the regime of low mass stars in 47~Tucanae, 
 ranging from $\sim$0.1~${\rm M_{\odot}}$, and reaching the H-burning limit and beyond.
First photometric data for this program, observed on July, 13, 2022 have recently demonstrated the $JWST$'s capability in uncovering multiple
 populations within the low-mass realm of M dwarfs marking indeed a pivotal advancement in the possibility of understanding the formation and evolution of GCs \citep[][]{milone2023a}.

Here, we present the latest observations collected in September 2023, including both photometric and spectroscopic data, which allowed us to explore the faintest stars of 47~Tucanae, encompassing a number of astrophysics issues, from the analysis of multiple stellar populations to the observation of brown dwarfs, and to the spectral detection of chemical variations.
For the first time, we provide indeed direct spectroscopic evidence of
multiple populations among M dwarfs, obtained from the first spectra ever observed in the faint MS of a GC. 

The overview of this work is as follows: Section~\ref{sec:data} describes the data set; Section~\ref{sec:cmd} presents the color-magnitude diagrams (CMDs) obtained from NIRCam images, and focuses on very low mass stars, including M-dwarfs and the brown dwarfs, which are poorly explored in GCs.
 This section includes the analysis of multiple populations among M-dwarfs, based on the CMDs. 
 We also present in Section\,\ref{sec:NIRSpec} early results based on the spectra of two M dwarfs associated with stellar populations with very different chemical composition, whereas Section\,\ref{sec:pratio} compares the fraction of 1P and 2P stars among stars with different masses. Finally, Section~\ref{sec:summary} discusses and summarizes our results.

\section{Data and data reduction}\label{sec:data}

To investigate low-mass stars in 47\,Tucanae with photometry, we used
images collected with the near infrared camera (NIRCam) on board 
$JWST$, the Wide-field Channel of the Advanced Camera (WFC/ACS), the
Ultraviolet and Visual channel and the Infrared Channel of the Wide
Field Camera 3 of $HST$ (UVIS/WFC3 and IR/WFC3). 
The dataset consists in NIRCam images of a field, denoted as C and
positioned approximately 11 arcminutes (corresponding to $\sim$3.5
  half light radii, assuming a half light radius of 3.17$\arcmin$
  from \citet{harris1996a}) westward from the cluster
center, (RA$\sim$00$^{\rm h}$21$^{\rm  m}$16$^{\rm s}$, DEC$\sim
-72^{\rm d}$06$^{\rm m}$16$^{\rm s}$), which is observed as part as
GO-2560 (PI A.\,F.\,Marino). 
We acquired images of field C simultaneously using the F115W filter of
the short-wavelength channel and the F322W2 filter of the
long-wavelength channel. These images were captured using the DEEP8
readout pattern and were dithered to effectively cover the gaps
between the $A$ and $B$ detectors of the short-wavelength channel.
In addition, we used data of a different field, denoted as A
(RA$\sim$00$^{\rm h}$22$^{\rm m}$37$^{\rm s}$, DEC$\sim -72^{\rm
  d}$04$^{\rm m}$06$^{\rm s}$) and located about 5~arcmin
  ($\sim$1.5 half light radii) westward
from the center of 47\,Tucanae. In particular, we analyzed three
regions of field A that have been observed as part of three distinct
visits of $JWST$ (GO-2559, PI I.\,Caiazzo). 

\begin{centering} 
\begin{figure} 
 \includegraphics[width=8.5cm,trim={0.5cm 0cm 0cm 0.0cm},clip]{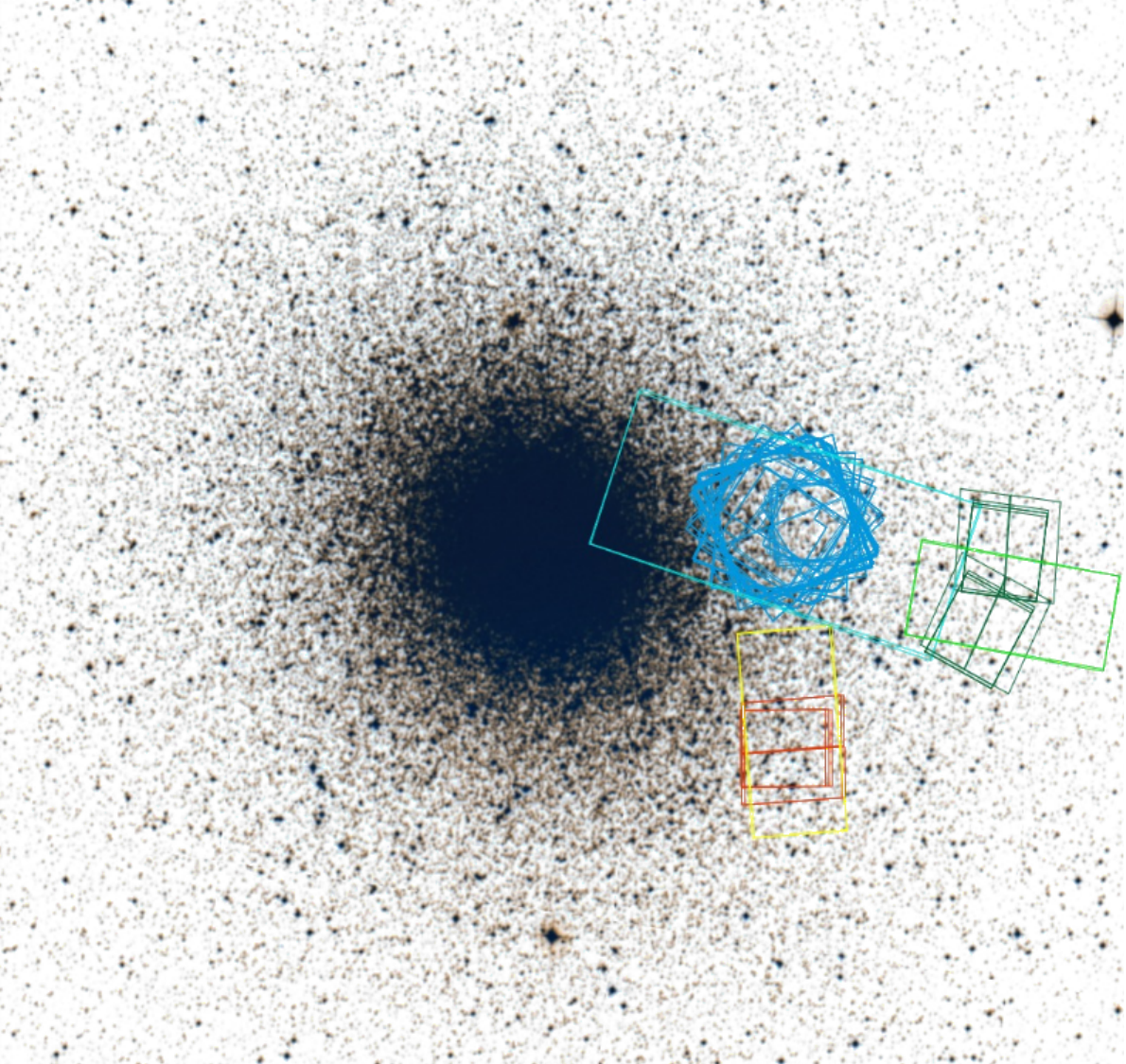}
 \caption{Footprints of the images used in our work on
   47\,Tucanae. Cyan and azure colors represent the $JWST$ and $HST$
   images, respectively, of field A, whereas the $JWST$ and $HST$
   footprints of field B are colored yellow and orange,
   respectively. Light and dark green colors indicate $JWST$ and $HST$
   images of field C. North is up, and east is left.}
 \label{fig:footprint} 
\end{figure} 
\end{centering}

The footprints of these images are shown in
Figure\,\ref{fig:footprint}, where we also show the images of Field B
(RA$\sim$00$^{\rm h}$22$^{\rm m}$36$^{\rm s}$, DEC$\sim -72^{\rm
  d}$09$^{\rm m}$27$^{\rm s}$) studied by \citet{milone2023a}. The
main properties of the $JWST$ images of fields A and C and the $HST$
images of field C are summarized in Table\,\ref{tab:data}, whereas we
refer to \citet{milone2023a} for details on the other images of fields
A and B\footnote{In addition to the NIRCam data listed in
    Table~\ref{tab:data}, field A comprises ACS/WFC images in 
  F606W and F814W, and IR WFC3 images in F105W, F110W, F140W, and
  F160W. Field B data consists in NIRCam F115W and F322W2 images collected as part of
  GO2560, and has been observed with $HST$ in F110W and F160W filters
  of the IR WFC3 camera and in F606W of UVIS/WFC3 \citep[see][for
  further details]{milone2023a}. }.
For completeness, we note that the line-of-sight based rotation
  axis of 47~Tucanae has inclination of $\sim$136$^{\circ}$ north to
  east \citep{bianchini2013, cordoni2020a}.

\begin{table*}
\caption{Description of the images used in the paper. For each dataset, we indicate the mission ($JWST$ or $HST$), the camera, the filter, the date, the exposure times, the GO program, and the principal investigator. We also indicate the visit numbers of GO-2559 observations. The observations collected during visit 1, 2, and 3 are centered around (RA,DEC)=(0$^{h}$ 22$^{m}$ 6.0$^{s}$, $-72^{d}$ 03$^{m}$ $51^{s}$), (0$^{h}$ 22$^{m}$ 13.9$^{s}$, $-72^{d}$ 05$^{m}$ $44^{s}$), and (0$^{h}$ 23$^{m}$ 11.2$^{s}$, $-72^{d}$ 04$^{m}$ $25^{s}$), respectively. }\label{tab:data}
\begin{tabular}{cccllll}
\hline \hline
 MISSION & CAMERA & FILTER  & DATE & N$\times$EXPTIME & GO & PI \\
\hline
         &          &          &     Field A                &                   &        & \\
  $JWST$   &  NIRCam  &  F150W2   &     September, 14, 2022  & 16$\times$857s   &  2559 Visit 1        & I.\,Caiazzo   \\
  $JWST$   &  NIRCam  &  F322W2   &     September, 14, 2022  & 16$\times$857s   &  2559 Visit 1        & I.\,Caiazzo   \\
  $JWST$   &  NIRCam  &  F150W2   &     September, 15, 2022  & 16$\times$857s   &  2559 Visit 2        & I.\,Caiazzo   \\
  $JWST$   &  NIRCam  &  F322W2   &     September, 15, 2022  & 16$\times$857s   &  2559 Visit 2        & I.\,Caiazzo   \\
  $JWST$   &  NIRCam  &  F150W2   &     September, 15, 2022  & 16$\times$857s   &  2559 Visit 3        & I.\,Caiazzo   \\
  $JWST$   &  NIRCam  &  F322W2   &     September, 15, 2022  & 16$\times$857s   &  2559 Visit 3        & I.\,Caiazzo   \\
         \hline
                  &          &          &     Field C                &                   &        & \\
  $JWST$   &  NIRCam  &  F115W   &     September, 24, 2023  & 38$\times$1031s   &  2560        & A.\,F.\,Marino   \\
  $JWST$   &  NIRCam  &  F322W2  &     September, 24, 2023  & 38$\times$1031s   &  2560        & A.\,F.\,Marino   \\
   $HST$   &  UVIS/WFC3 &  F606W  &    April, 10, 2010 & 2$\times$50s$+$1347s$+$1398s   &  11677 &  H.\,B.\,Richer  \\
   $HST$   &    IR/WFC3 &  F110W  &      April, 10, 2010 & 102s$+$174s$+$2$\times$1399s   &  11677 &  H.\,B.\,Richer  \\
   $HST$   &    IR/WFC3 &  F160W  &      April, 10, 2010 & 4$\times$299s$+$4$\times$1199s &  11677 &  H.\,B.\,Richer  \\
     \hline\hline
     \end{tabular}
\end{table*}

To measure stellar fluxes and positions we used a method that is based on two main steps.
First-step photometry and astrometry is obtained by using the computer program {\it img2xym}.
Originally devised by Jay Anderson \citep[e.g.\,][]{anderson2006a} for
the reduction of $HST$ images, the methodology involves independently
measuring stellar fluxes and positions in each image. This is achieved
by employing a spatially variable point-spread-function (PSF) model
\citep{anderson2000a} along with a 'perturbation PSF' that fine-tunes
the fitting process to accommodate slight variations in the telescope
focus. 
The perturbation PSF is obtained through unsaturated, bright, and isolated stars. 
All magnitude determinations derived from individual filters and
cameras have been standardized to a common photometric zero point,
aligning with the zero point of the deepest exposure in the chosen
filter, which serves as the reference frame for constructing the
photometric master frame. This alignment was achieved by utilizing
bright, unsaturated stars well-fitted by the PSF to calculate the
magnitude differences between the master frame and each exposure. The
mean of these differences was then used to transform star measurements
in each exposure into the reference frame. 
For geometric distortion correction, solutions provided by
\citet{Anderson2022a}, \citet{bellini2011a}, and \citet{milone2023a}
were applied to adjust stellar positions of UVIS/IR, UVIS/WFC3, and
NIRCam images. The coordinates of stars in all cluster images were
transformed into a common reference system based on Gaia Data Release
3 (DR3) catalogs \citep{gaia2021a}. This transformation ensures
alignment with the West and North directions for the abscissa and
ordinate, respectively. 

The second-step photometry and astrometry of all sources has been
carried out with the computer program KS2, which is developed by Jay
Anderson and is based on the program {\it kitchen\_sync} by
\citet{anderson2008a} \citep[see][for details]{sabbi2016a,
  bellini2017a, milone2023b}.  

KS2 employs three distinct methods for measuring stars, and each
method provides optimal photometry for stars with varying levels of
brightness.  
In method I, which works well for bright stars,  the stellar fluxes
and positions were independently derived in each individual exposure
by using the most suitable effective point-spread function (PSF) model
for their specific location on the detector.  
The sky brightness is measured over an annular region between 4 and 8
pixels from the center of the star. 

Method II combines information from all exposures to derive the magnitudes of stars in each exposure by means of aperture photometry, after subtracting the nearby stars. 
 Specifically, we used a 5$\times$5 pixel aperture and determined the sky as in method I.
This method works well for faint stars, which often do not have enough photons to constrain their fluxes and positions in the individual exposures. Method III is similar to method II. The main difference is that the aperture photometry is computed by considering a circular region with a radius of 0.75 pixels, and the background sky is determined from the annulus located between 2 and 4 pixels away from the initially identified position.
Subsequently, the multiple measurements for each star were averaged to obtain the most accurate estimations of their magnitudes and positions.

We used the parameters that are indicative of the astrometric and photometric quality provided by the KS2 computer programs to select the stars that are well-fitted by the PSF model. To do this, we used the procedure by \citet[][see their section 2.4]{milone2023b}.
Photometry has been calibrated to the Vega system as in \citet{milone2023b} and by using the zero points available in the Space Telescope Science Institute webpage\footnote{https://www.stsci.edu/hst/instrumentation/wfc3/data-analysis/photometric-calibration; https:\,//jwst-docs.stsci.edu\,/jwst-near-infrared-camera\,/nircam-performance\,/nircam-absolute-flux-calibration-and-zeropoints} for WFC3 and NIRCam. We verified that the photometry is not affected by significant reddening variations across the field of view \citep{legnardi2023a}. Hence, we did not correct the photometry for differential reddening.

\begin{figure*} 
\centering
 \includegraphics[width=8.2cm,trim={1.0cm 4cm 0.0cm 3.0cm},clip]{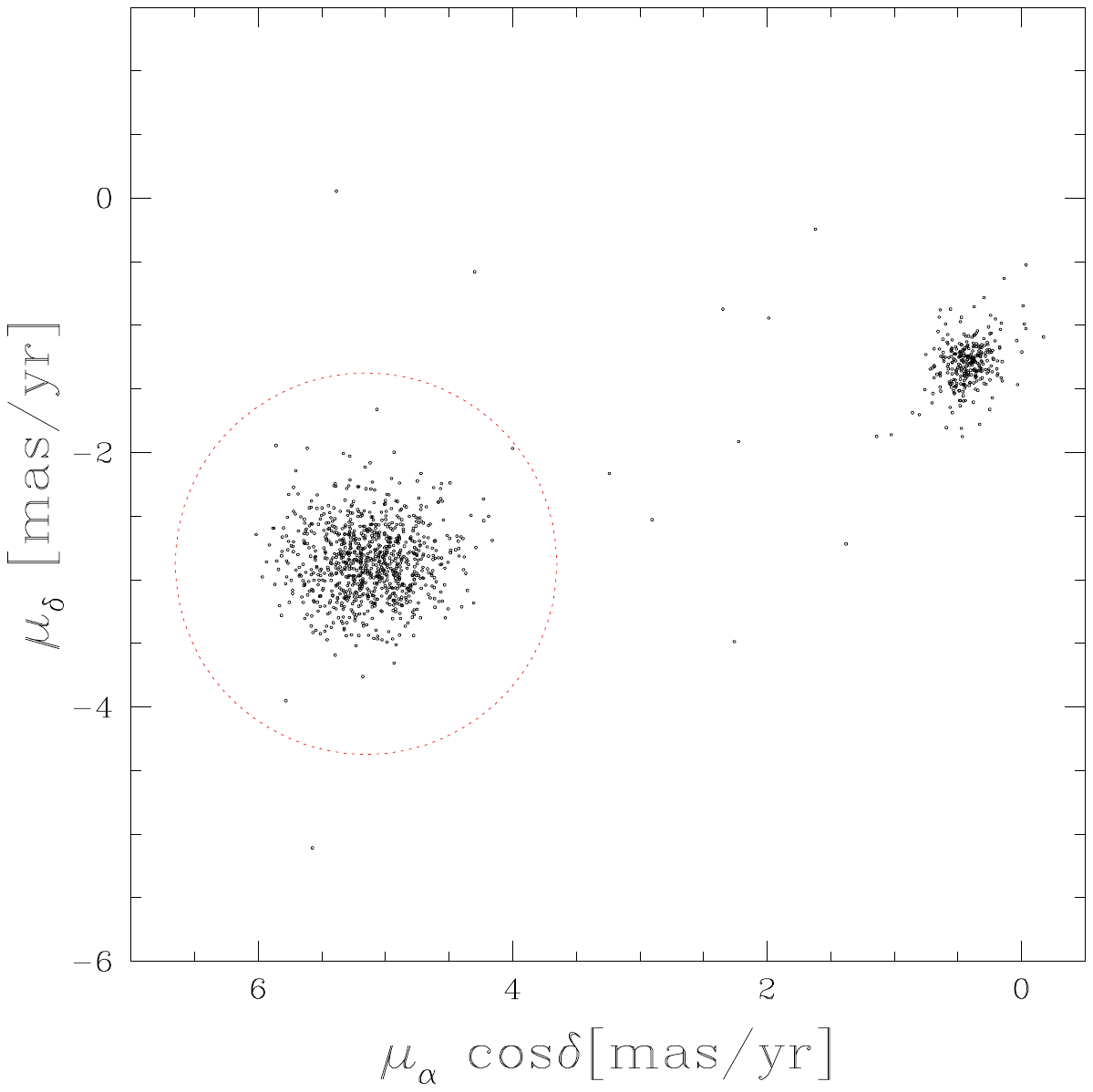} 
  \includegraphics[width=8.2cm,trim={1.0cm 4cm 0cm 3.0cm},clip]{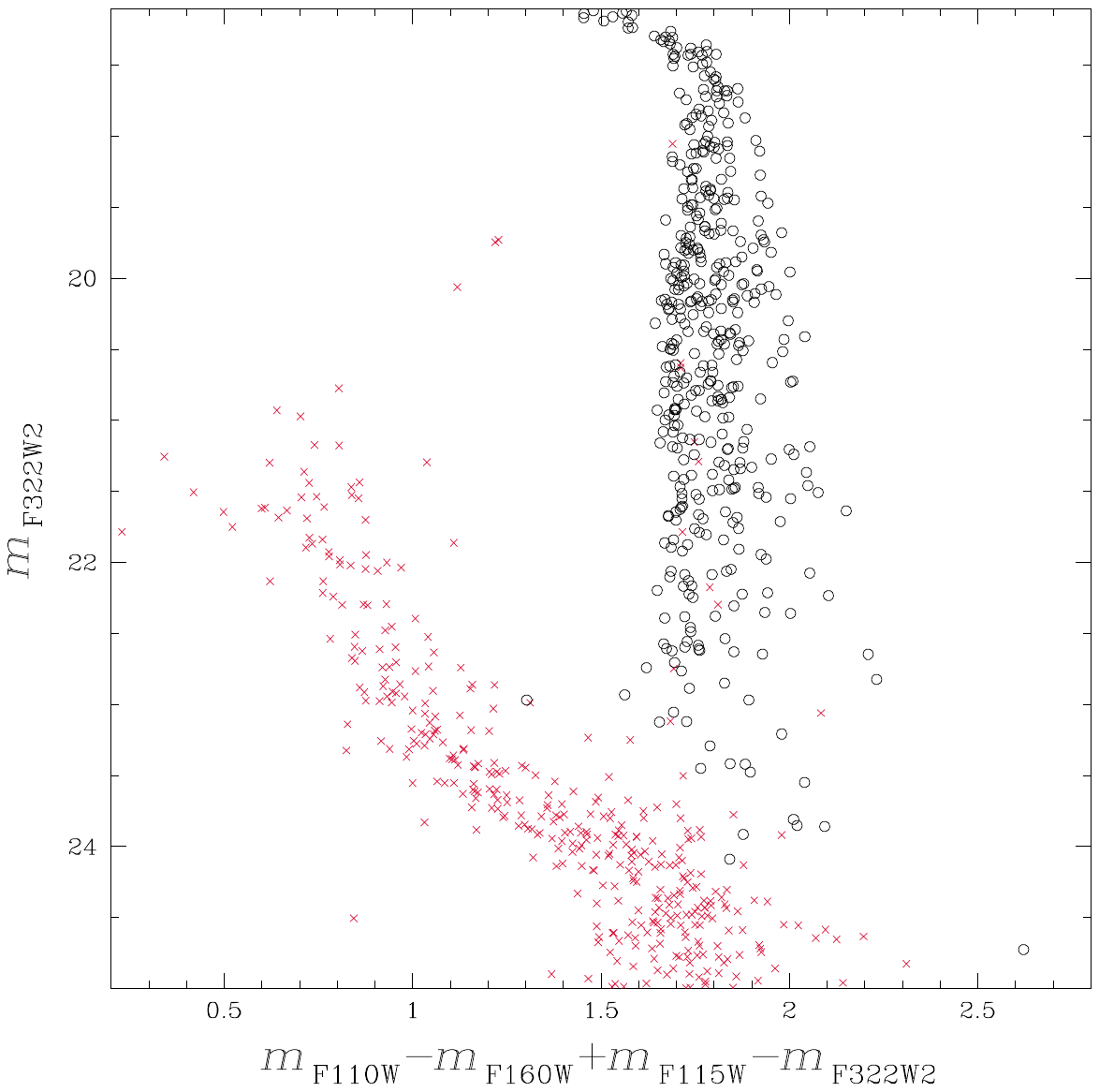}
 \caption{
 {\it Left.} Proper motions for stars in field C brighter than $m_{\rm
   F322W2}=24.0$ mag. Stars in 47\,Tucanae, within the red dotted
   circle, and SMC are clearly separated in this plot. {\it Right.}
   $m_{\rm F322W2}$ vs.\,$m_{\rm F110W}-m_{\rm F160W}+m_{\rm
     F115W}-m_{\rm F322W2}$ pseudo-CMD for probable cluster members
   (black circles), which are located within the red dotted circle in
   the left-panel diagram, and for SMC stars (red crosses).} 
 \label{fig:PMs} 
\end{figure*}

Proper motions are derived as in \citet[][see their section
2.3]{milone2023a} by comparing the distortion-corrected positions of
stars measured at different epochs. 
The proper-motion diagram derived for field-C stars brighter than $m_{\rm F322W2}=24.0$ is plotted in the left panel of Figure\,\ref{fig:PMs}, showcasing two primary stellar clumps encompassing the majority of cluster members and Small Magellanic Cloud (SMC) stars. The red circle is used to distinguish probable 47~Tucanae members from field stars, which are colored black and red, respectively, in the $m_{\rm F322W2}$ vs.\,$m_{\rm F110W}-m_{\rm F160W}+m_{\rm F115W}-m_{\rm F322W2}$ pseudo-CMD.

\begin{figure*} 
\centering
 \includegraphics[height=10.cm,trim={0.5cm 5.1cm 4.5cm 9.5cm},clip]{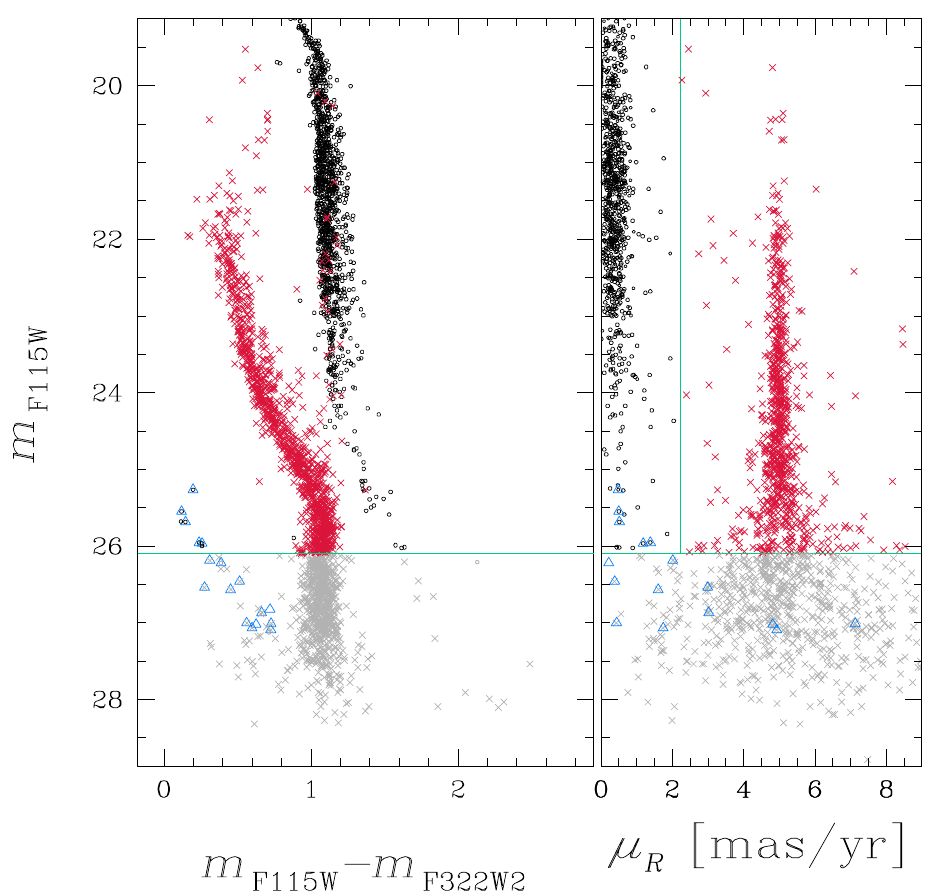} 
  \includegraphics[height=10.cm,trim={1.0cm 5.1cm 10cm 9.5cm},clip]{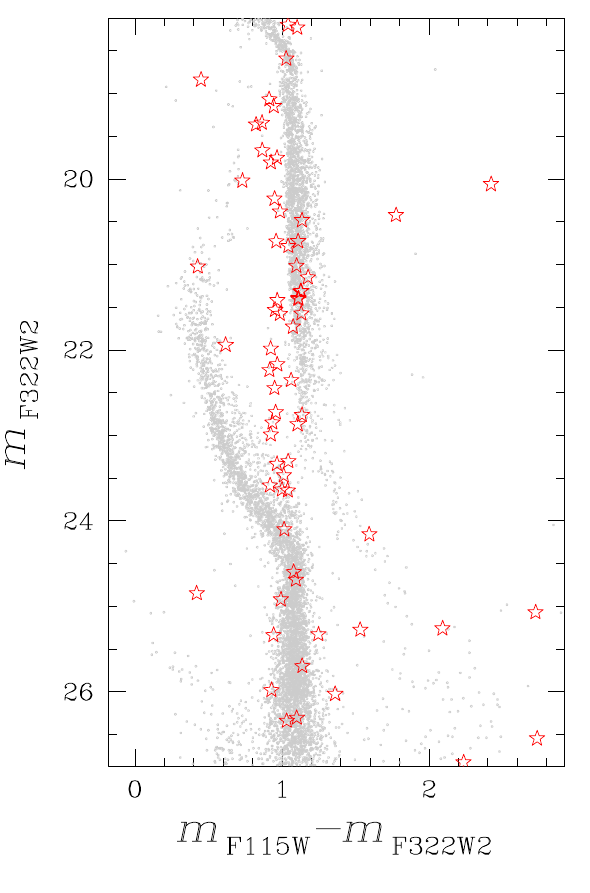}
 \caption{
 $m_{\rm F115W}$ vs.\,$m_{\rm F115W}-m_{\rm F322W2}$ CMD for stars in the field C with available proper motions (left panel). The middle panel shows the F115W magnitude as a function of the proper motion relative to 47\,Tucanae. The vertical aqua line separates the bulk of cluster members (black points in the left and middle panel) from the field stars (red crosses). The stars below the aqua horizontal line, where the proper motions do not allow us to disentangle field stars from cluster members, are represented with grey crosses. Probable white dwarfs, selected from the $m_{\rm F606W}$ vs.\,$m_{\rm F606W}-m_{\rm F115W}$ CMD are represented with blue  triangles in the left and middle panels. In the right panel we show the $m_{\rm F322W2}$ vs.\,$m_{\rm F115W}-m_{\rm F322W2}$ CMD for all stars in the field CMD (grey points) with the stars simulated by the trilegal Galactic model \citep[red starred symbols,][]{girardi2005a} for a field with the same area as field C in the direction of 47\,Tucanae.}
 \label{fig:PMs2} 
\end{figure*}

The $m_{\rm F115W}$ vs.\,$m_{\rm F115W}-m_{\rm F322W2}$ CMD of all stars with available proper motions is plotted in the left panel of Figure\,\ref{fig:PMs2}. We computed for each star the proper motion relative to the average proper motion of 47\,Tucanae, $\mu_{\rm R}$, and plotted this quantity against the F115W magnitude.
For magnitudes brighter than $m_{\rm F115W}=26.1$ mag the stars of 47\,Tucanae and the SMC stars exhibit distinct proper motions. Hence, we draw by eye the vertical aqua line to separate the bulk of cluster members from the field stars, which are colored black and red, respectively in the left and middle panels of Figure\,\ref{fig:PMs}. 
 Due to the large observational errors, the proper motions of stars fainter than $m_{\rm F115W}=26.1$ mag do not allow us to disentangle field stars from cluster members.

 To estimate the amount of faint field stars that we expect in the field C, we compare the observed $m_{\rm F322W2}$ vs.\,$m_{\rm F115W}-m_{\rm F322W2}$ CMD of all stars with available NIRCam photometry and the simulated CMD calculated by the Trilegal code \citep{girardi2005a} for a Galactic field with the same area and the same Galactic coordinates as field C. The result is plotted in the right panel of Figure\,\ref{fig:PMs2} where we observe that a negligible fraction of Milky Way field star
 interlopers are expected to contaminate the observed CMD sequences of 47\,Tucanae.

\subsection{Artificial stars and completeness}

We conducted artificial star (AS) experiments to estimate the photometric uncertainties and the completeness level of our sample by using the procedure adopted in previous papers \citep[e.g.][]{anderson2008a}. 
These stars were distributed across the field of view similarly to the cluster stars and along the fiducial lines of the main sequence (MS) of 47\,Tucanae that we derived from the observed CMDs. The ASs have been reduced by using exactly the same procedure adopted for the real stars.
The KS2 computer program generates the same diagnostic measurements of photometric and astrometric quality for ASs as it does for real stars. In our analysis, we considered only a subset of artificial stars that are relatively isolated, exhibit good PSF fitting, and show small root mean square (rms) values in magnitudes. These artificial stars were selected using the same criteria we applied to real stars in our investigation.

We partitioned the NIRCam field for each field into five concentric anuli, centered on 47\,Tucanae. Within each annulus, we analyzed the results of the artificial star experiments in a number of $N$ distinct 0.5 magnitude bins, spanning from the saturation limit to approximately one F322W2 magnitude below the faintest star that we detected.
For each of the 5$\times N$ grid positions, we calculated the average completeness by comparing the number of recovered ASs with the input artificial stars within that particular bin.

\subsection{NIRSpec spectra}\label{subsec:spectra}

NIRSpec has been used in the multi-object spectroscopy (MOS) mode, involving a micro-shutter assembly (MSA) configuration, which allowed us the simultaneous collection of 29 source spectra of sufficient quality, within a $3.4' \times 3.6'$ field of view centered on field B.
We choose the G235M/F170LP disperser and filter combination, which observes the wavelength range 1.66-3.07~$\mu$m, at a nominal resolving power of $\sim$1,000.

The improved reference sampling and subtraction (IRS$^{2}$) readout mode, with the NRSIRS2 pattern, which has 5 frames averaged into a single group, has been employed for our MOS observations.
Each source has been observed with a 2-Shutter nod in Slitlet pattern with two identical configurations and each configuration has been executed ten times. 
Each exposure was observed for 1,182s (with 16 groups per integration), for a total on-target time of 47,280s.
However, a dither of 20$\arcsec$ in the dispersion direction has been applied to close the detector gap.
In the end, of our 29 observed stars, 17 are covered by both dither positions, hence have all the 40 exposures, while 12 have only one dither position (20 exposures). 

We have analysed the NIRSpec 1-D extracted spectra processed with the $JWST$ Science Calibration Pipeline \citep{Bushouse2023}.
While for a full analysis of the individual stars we refer to a forthcoming paper devoted to the oxygen estimates, we anticipate in this paper the results for spectra of M-dwarfs with extreme chemical compositions.

\section{The NIRCam CMD of 47\,Tucanae}\label{sec:cmd}

The $m_{\rm F322W2}$ vs.\,$m_{\rm F115W}-m_{\rm F322W2}$ CMD of stars
in the field C is illustrated in the lower panels of
Figure\,\ref{fig:CMDs}, whereas the upper panels show portions of the
NIRCam field of view. 
We notice that the $m_{\rm F115W}-m_{\rm F322W2}$ color broadening of
the upper MS of 47\,Tucanae ($m_{\rm F322W2} \lesssim 18.7$) is
comparable with the broadening due to the observational errors alone.  
In contrast, the MS color spread for stars fainter than the MS knee is much wider than the observational errors.
This phenomenon, which is due to the presence of multiple populations
among the M-dwarfs, will be further investigated in
Section\,\ref{sec:MPs}. 
In the following Section\,\ref{sec:verylowmass} we investigate the
stars at faintest magnitudes approaching the H-burning limit. 

\subsection{Ultracool dwarfs}\label{sec:verylowmass}

\begin{figure*} 
 \includegraphics[width=15.5cm,trim={1.3cm 5.5cm 1.0cm 0.0cm},clip]{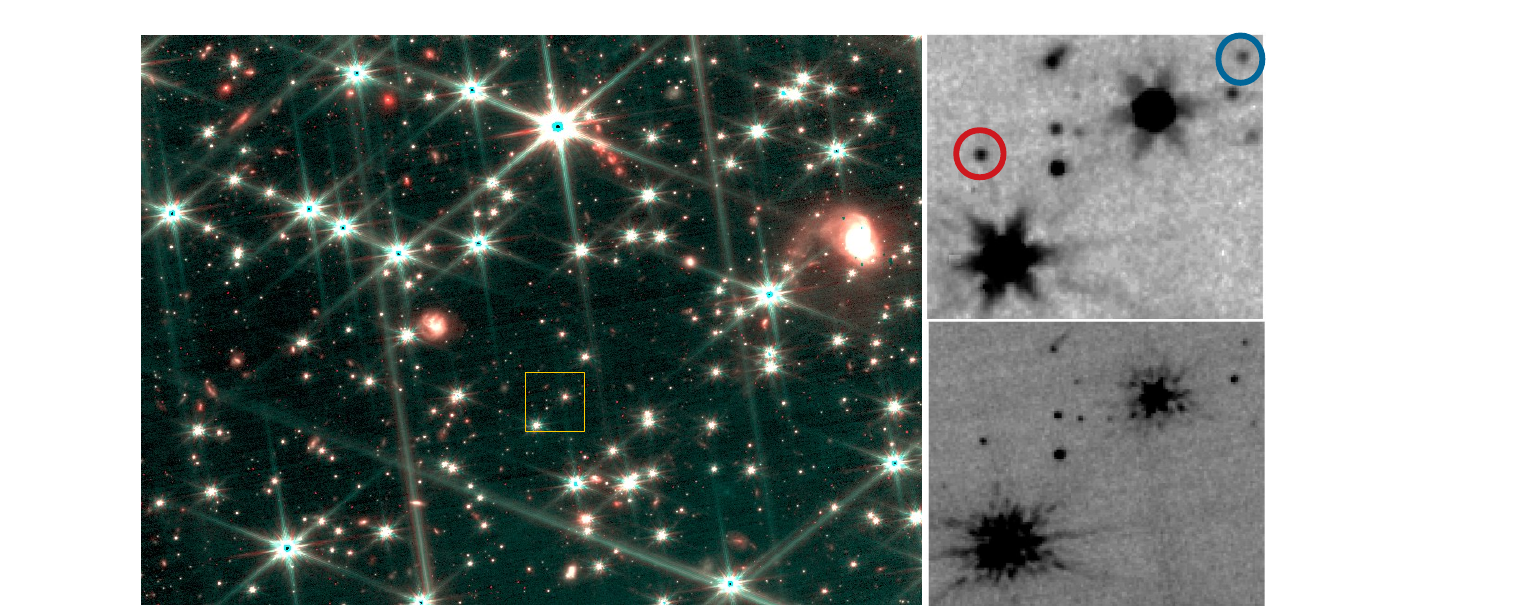} \\
  \includegraphics[width=13.0cm,trim={0.5cm 5.5cm 0cm 9.0cm},clip]{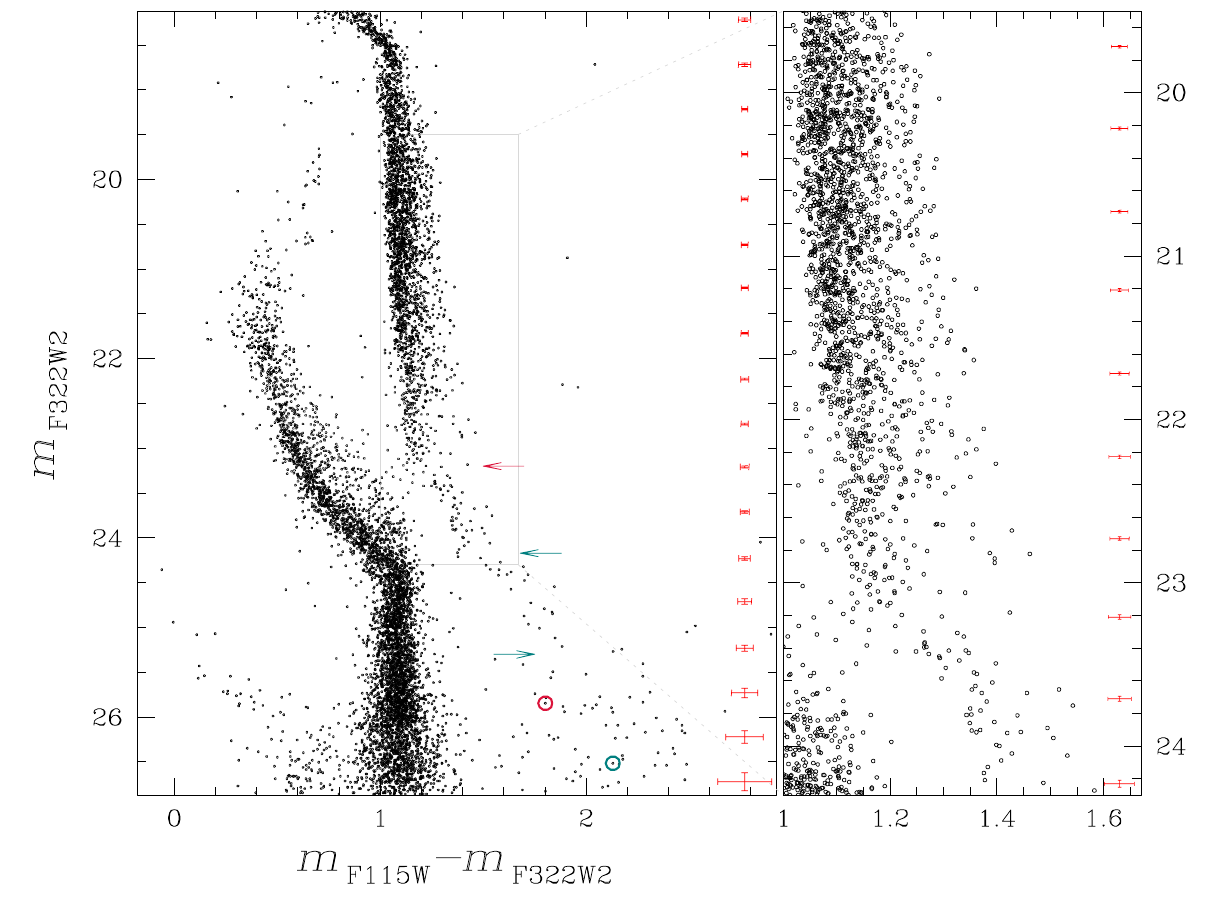}
 \caption{
 {\it Top.} Three-color (RGB) image of a portion of the studied field of view (field C). We used the stacked F115W and F322W2 images for the B and R channels, respectively, whereas the G channel is obtained by averaging the F115W and F322W2 stacked images (left). The right panels show the F322W2 (top) and F115W (bottom) stacked images for the region within the yellow box highlighted in the field of view on the left panel.
 {\it Bottom.}  
 $m_{\rm F322W2}$ vs.\,$m_{\rm F115W}-m_{\rm F322W2}$ CMD of stars in field C (left). The arrows indicate the location of the possible gaps or discontinuities along the sequence of ultracool 47\,Tucanae stars.
  Stars fainter than $m_{\rm F322W2} \sim 25.3$ mag and with colors redder than $m_{\rm F115W}-m_{\rm F322W2} \sim 1.5$~mag are probable brown dwarfs. 
  The blue and red circles in the CMD highlight two of them, which are also marked in the top-right panel. We choose these two stars, among the brown dwarf sample, for illustrative purposes, as they are located in a relatively small region of the field of view.   
  A zoom around the MS region of 47\,Tucanae populated by M-dwarfs is shown in the right panel.}
 \label{fig:CMDs} 
\end{figure*}

\begin{figure*} 
\centering
\includegraphics[width=14.5cm,trim={0.5cm 5.5cm 0.2cm 4.5cm},clip]{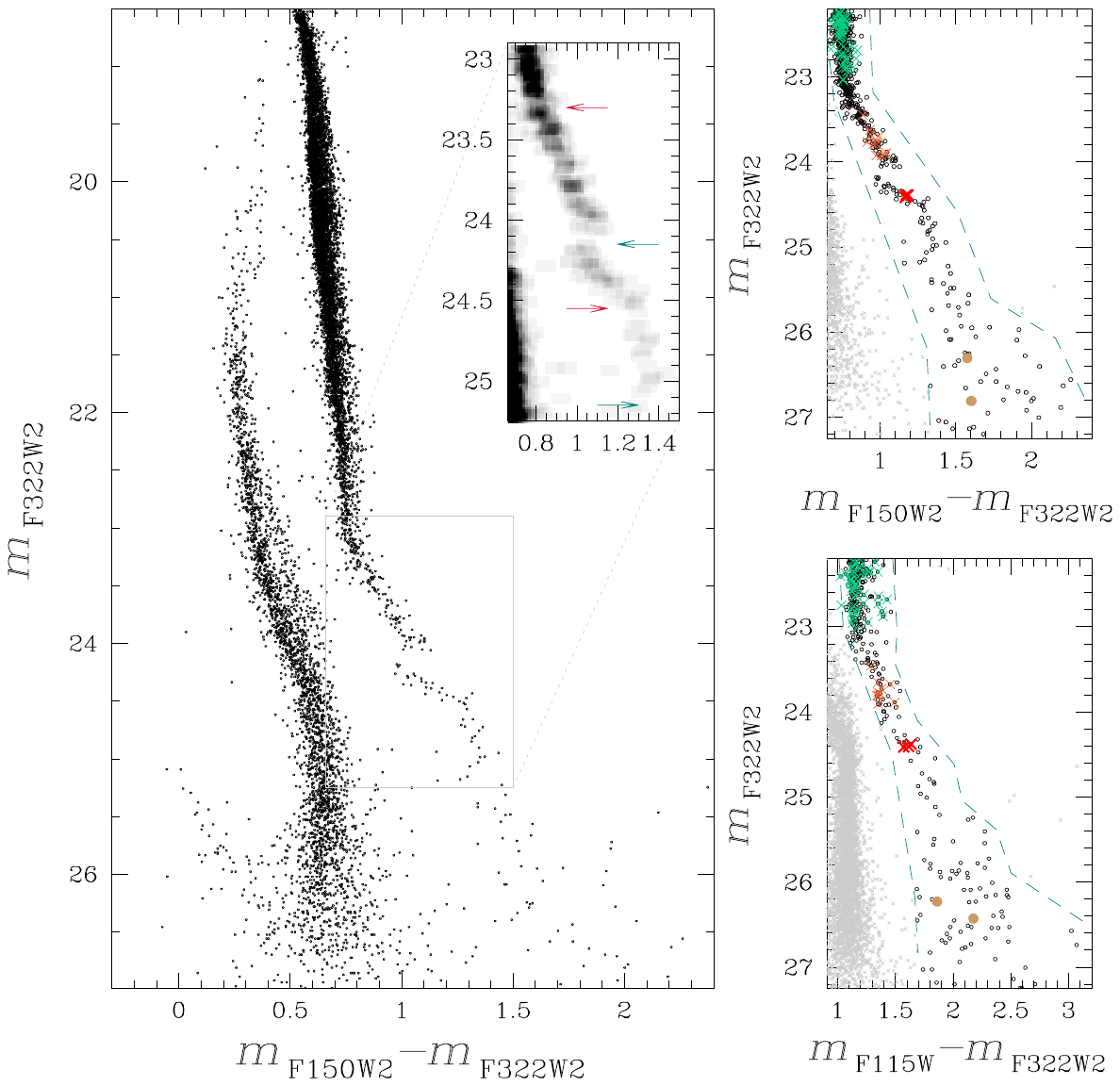}
 \caption{{\it Left}. $m_{\rm F322W2}$ vs.\,$m_{\rm F150W2}-m_{\rm
     F322W2}$ CMD of stars in field A (left). The inset shows the Hess
   diagram of the low MS of 47\,Tucanae, where the main gaps and
   discontinuities are marked with colored arrows. Specifically
     the crimson arrows indicate the changes in slope at $m_{\rm
       F322W2}$ around 23.3 and 24.5~mag, while the teal arrows
     correspond to the sharp gap at $m_{\rm F322W2}\sim$24.2, and the
     observed minimum in the luminosity function at $m_{\rm
       F322W2}\sim$25.3~mag. {\it Right}. $m_{\rm F322W2}$
     vs.\,$m_{\rm F150W2}-m_{\rm F322W2}$  (top) and $m_{\rm F322W2}$
     vs.\,$m_{\rm F115W}-m_{\rm F322W2}$ CMDs (bottom) for faint MS
     stars and brown-dwarfs. The stars observed in both CMDs are
     represented with colored symbols, specifically: 
 the aqua, orange, and red colors indicate the dwarfs that populate
 the main MS segments, whereas the likely brown dwarfs observed in
 both datasets are colored brown. } 
 \label{fig:CMDs2} 
\end{figure*}

We analyse here the faintest 47\,Tucanae stars appearing in the
$m_{\rm F322W2}$ vs.\,$m_{\rm F115W}-m_{\rm F322W2}$ CMD of
Figure\,\ref{fig:CMDs}. In particular, we focus on the faintest
portion of the CMD, composed by a plume of stars extending from the
bottom of the vertical sequence (at $m_{\rm F322W2} \sim 23.3$~mag)
down to $m_{\rm F322W2} \sim 27.0$ mag. This sparsely populated group
of stars covers a wide color range ($\sim$2 mag) and it appears as an
extension of the bluest portion of the vertical sequence composed by
the cluster's M dwarfs. 
A quick glance at this sequence suggests that its stars, which have
masses less than $\sim 0.1$ M$_{\odot}$, are not continuously
distributed. In particular, we notice the presence of compelling
features, namely, possible clumps at $m_{\rm F322W2} \sim 23.3$, a
sharp gap $\sim 24.2$~mag, and a distinct change in slope at $\sim
24.5$~mag (see also Figure\,\ref{fig:CMDs} and
Figure\,\ref{fig:CMDs2}). 

The observed luminosity function  approaches its minimum value around
$m_{\rm F322W2} \sim 25.3$,  and rises up toward fainter luminosities,
where we observe the bright portion of the brown-dwarf sequence. As an
example of the analysed field of view, the upper-right panels of
Figure\,\ref{fig:CMDs} show a zoom of the stacked F322W2 and F115W
images that include two probable brown dwarfs (red and blue circles).

To further investigate the lower MS of 47\,Tucanae, we show in the left panel of Figure\,\ref{fig:CMDs2} the $m_{\rm F150W2}$ vs.\,$m_{\rm F150W2}-m_{\rm F322W2}$ CMD of stars in Field A. As highlighted by the Hess diagram plotted in the inset, we confirm the main {gap and bend in the CMD already evident in Figure\,\ref{fig:CMDs}.  
 In particular, this diagram clearly shows the change in the slope of the sequence of ultracool stars at} $m_{\rm F322W2}=24.5$~mag.
The right panels of Figure\,\ref{fig:CMDs2} compare the $m_{\rm
  F322W2}$ vs.\,$m_{\rm F150W2}-m_{\rm F322W2}$  (top) and the $m_{\rm
  F322W2}$ vs.\,$m_{\rm F115W}-m_{\rm F322W2}$ CMDs (bottom) zoomed
around the bottom of the MS, that are derived from GO-2559 and GO-2560 data. 
The dashed lines, derived by eye, enclose the bulk of stars along the MS and the brown-dwarf sequence that we marked with black circles. The colored symbols mark the stars for which photometry from both GO-2559 and GO-2560 is available.

\begin{figure*} 
\centering
\includegraphics[height=11.0cm,trim={0.5cm 5.2cm 9cm 4.0cm},clip]{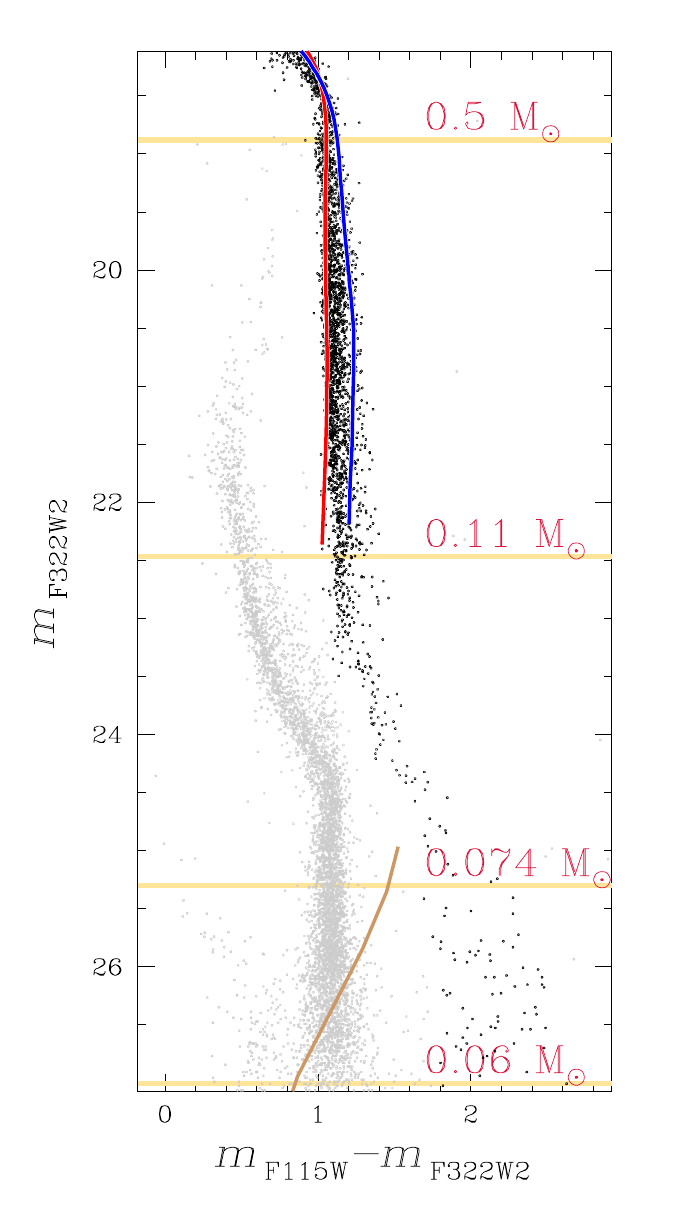}
\includegraphics[height=10.7cm,trim={6.0cm 5.2cm 0cm 4.0cm},clip]{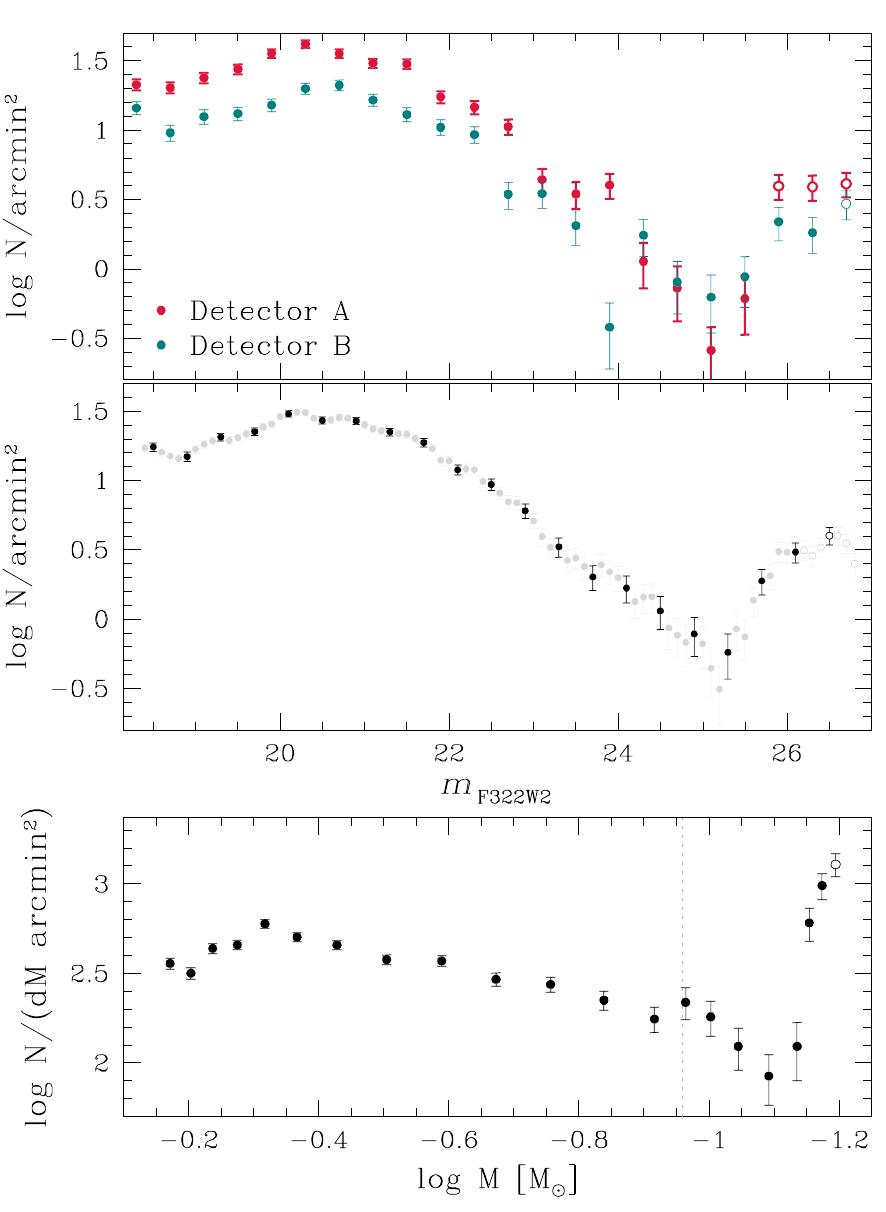}
 \caption{{\it Left.} Reproduction of the $m_{\rm F322W2}$
   vs.\,$m_{\rm F115W}-m_{\rm F322W2}$ CMD of Figure\,\ref{fig:CMDs},
   where we used black colors to mark the probable MS and brown dwarfs
   of 47\,Tucanae. We superimposed on the CMD the best-fit isochrones
   for MS stars \citep{dotter2008a, milone2023a} and the isochrone by
   \citet{phillips2020a} for low-mas stars. The horizontal lines mark
   the magnitudes corresponding to four stellar masses inferred from
   the red isochrone and from \citet{gerasimov2023a}. {\it Right.}
   F322W2 luminosity function of 47\,Tucanae stars along the MS and
   the brown-dwarf sequence obtained by using 0.4-wide magnitude
   bins. The open circles are derived from stars where the correction
   for the incompleteness of the photometric sample is smaller than
   50\%. 
  The top panel shows the luminosity functions of stars in the
  detectors A and B of the long-wavelength channel of NIRCam,
  separately, which have different radial distances of $\sim$11.5 and
  14.5 arcmin from the cluster center. The middle panel shows the
  results from the entire field of view.  The grey points in the
  middle panel are derived by using 0.4-wide magnitude bins but by
  changing the starting magnitude value of the LF by 0.1, 0.2, and 0.3
  mag with respect to the value used for the black points. The bottom
  panel shows the mass function for stars in the entire field of
  view. The vertical dotted line corresponds to 
0.11 solar masses. See the text for details.}
 \label{fig:LFs} 
\end{figure*}

The luminosity and mass functions may provide additional information
of low mass stars in 47\,Tucanae. Unfortunately,  isochrones that
reproduce both massive and low-mass stars in 47\,Tucanae are not
available to us as shown in the left panel of Figure\,\ref{fig:LFs},
where we compare the $m_{\rm F322W2}$ vs.\,$m_{\rm F115W}-m_{\rm
  F322W2}$ CMD with three isochrones. We have marked with black dots
the probable cluster MS and brown dwarfs that we selected by eye,
whereas the white dwarfs of 47\,Tucanae and the Small Magellanic Cloud
(SMC) stars are colored grey. 
The red and the blue isochrones  provide the best fit of 1P and 2P
stars with extreme chemical composition, respectively, and comprise
stars more massive than $\sim 0.1\, {\rm M}_\odot$. These isochrones,
which refer specifically to 47\,Tucanae and are described in detail by
\citet{milone2023a}, are extracted  from the Dartmouth database
\citep{dotter2008a} and share the same age (13~Gyr), iron abundance
([Fe/H]=$-$0.75) and [$\alpha$/Fe]=$+$0.4~dex. However, they have
different abundances of He, C, N, and O, in such a way that the blue
isochrone is enhanced in helium mass fraction by $\Delta Y$=0.04 and
in [N/Fe] by 1.2~dex, and depleted in [C/Fe] and [O/Fe] by $-0.35$ and
0.50~dex, respectively, relative to the red isochrone that has
Y=0.248, [O/Fe]=0.4 and solar carbon and nitrogen abundances. 

To derive the mass-luminosity relation  for stars less massive
than $\sim 0.1\, {\rm M}_\odot$ 
we tentatively used the tracks by \citet{phillips2020a} (brown
isochrone plotted in the left panel of Figure\,\ref{fig:LFs}), which
however are based on model atmospheres and evolution models for brown
dwarfs and giant exoplanets with solar metallicity. Clearly, this
isochrone, which is computed for an age of 10~Gyr, does not match the
brown-dwarf sequence of 47\,Tucanae, and differs
from those derived by \citet{gerasimov2023a} that qualitatively reproduce the observed CMDs.
In a recent paper, \citet[][]{nardiello2023a} identified ten candidate
brown dwarfs by using the $m_{\rm F322W2}$ vs.\,$m_{\rm F115W}-m_{\rm
  F322W2}$ CMD of field-B stars in 47\,Tucanae. These stars,  are
distributed along the isochrone by \citet{phillips2020a}  and span
F322W2 magnitudes between $\sim$25.2 and 25.8~mag are superimposed on
the SMC MS and may then belong to SMC rather than to
47\,Tucanae. Accurate stellar proper motions, which are mandatory to
disentangle SMC stars and possible cluster members, are not available
for these very faint stars. Hence, we neither exclude nor confirm the
presence of brown dwarfs in 47\,Tucanae following the Phillips et
al. isochrone. In the following, we  only focus on the sequence of
ultracool stars analyzed in this paper. 

The luminosity functions of 47\,Tucanae stars are plotted in the
top-right and middle-right panel of Figure\,\ref{fig:LFs}, for the
stars observed with the detectors A and B of the long NIRCam channels
and for all stars together. Specifically, we plotted the logarithm of
the numbers of stars per square arcmin, corrected for completeness, as
a function of the $m_{\rm F322W2}$ magnitude. The star counts are
calculated in 0.4-mag wide bins. 
The luminosity function exhibits a peak around $m_{\rm F322W2} \sim
20.2$. As we move towards fainter luminosities, the star count per
unit magnitude consistently decreases,
reaching its minimum around $m_{\rm F322W2} \sim 25.3$~mag. 
Beyond this point, the number of stars increases for fainter values of $m_{\rm F322W2}$.

The bottom-right panel of Figure\,\ref{fig:LFs} shows a tentative
determination of the mass function, which is derived 
from the luminosity function plotted in the middle-right panel. For
simplicity, we did not account for the multiple populations in
47\,Tucanae, and we used the red isochrone plotted in the left panel
to convert the luminosities into stellar masses for stars brighter
than $m_{\rm F322W2} \sim 22.5$, which have masses larger than $\sim
0.1$~M$_{\odot}$,  In this mass interval, we find that the mass
function exhibits a peak at $\sim$0.5~M$_{\odot}$ and declines toward
lower masses. 

Due to the lack of appropriate isochrones for stars with masses lower
than $\sim0.1$~M$_{\odot}$, it is not possible to provide accurate
mass functions for these stars. Nevertheless,  
 we assumed that the luminosity  $m_{\rm F322W2} = 25.3$
  corresponds to a stellar mass of 0.074~${\rm M}_\odot$, which
  according to \cite{gerasimov2023a} corresponds to the
  hydrogen-burning limit, whereas stars with $m_{\rm F322W2} = 27$
  have 0.06 solar masses, as predicted by Gerasimov and
  collaborators. 
The results are plotted in the bottom-right panel of
Figure\,\ref{fig:LFs}, showing an abrupt change in the mass
  function slope  at the assumed hydrogen-burning limit. We are
  reluctant to take for good this result, because of the tentative
  mass-luminosity relation we have adopted.

\subsection{The M-Dwarf to Brown-Dwarf Transition in 47\,Tucanae.}

As emphasised by \cite{gerasimov2023a}, the termination of the hydrogen-burning MS is expected to exhibit a gap in the CMD between the faintest MS stars and the brightest brown dwarfs. Thus, it is tantalising to identify this gap (i.e., the termination of the MS) with the evident gap at $m_{\rm F322W2} \sim 24.2$ seen in Figure\,\ref{fig:CMDs2}. In their simulations Gerasimov et al. do not find  a similar gap at this luminosity (their Figure 6), whereas a gap may appear at a much fainter luminosity, $m_{\rm F322W2} \sim 26$ (their Figure 10). This corresponds to the minimum in their synthetic luminosity function at this latter magnitude, as illustrated in their Figure 9. Thus, \citet{[\,][]{gerasimov2023a}} associate this minimum with the hydrogen-burning limit. 

Our observed luminosity function shows  also a sharp minimum (see Figure\,\ref{fig:LFs}), but it is found at $m_{\rm F322W2} \sim 25.3$, i.e., around half magnitude brighter than predicted by Gerasimov et al.
So, where is the hydrogen-burning limit in 47 Tuc? Is it at $m_{\rm F322W2} \sim$ 24.2, at 25.3, or at 26? 
What would be needed is an extended set of isochrones, possibly with options in the input physics. The data we present here represents a major jump in depth and completeness, for coeval M and brown dwarfs all at the same distance and extending down to unprecedented faint luminosities. These data demand a new, parallel effort in modelling M dwarfs and brown dwarfs.

In the meantime, we can still speculate on the nature of of the gap at $m_{\rm F322W2} \sim 24.2$ and the origin of the different luminosities at the minimum of the luminosity function, just mentioned above.
If the gap does not correspond to the hydrogen-burning limit, what
else can cause it? These M dwarfs are fully convective,  so convection
effects could hardly result in such a sharp discontinuity (note that a
drop is not evident in the luminosity function at this magnitude
because of the binning, except in the case of blue points in
Figure\,\ref{fig:LFs}). Could atmospheric effects cause it? We don't
know, so it remains an unsolved issue. 

Similarly, we cannot unambiguously pinpoint the origin of the
mag discrepancy in the minimum of the luminosity function. Both sets
use Vega magnitudes, so it cannot be a zero-point issue. Maybe it is
due to a systematic effect on the stellar models, in a regime in which
a small difference in mass can result in a large difference in
luminosity. On the observational side, at these faint magnitudes the
luminosity function of galaxies is raising steeply, so we cannot
exclude that our luminosity function of 47 Tucanae is partly
contaminated by faint galaxies. Still we have excluded morphologically
non-stellar objects in constructing the luminosity functions of 47\,Tucanae.

\subsection{Multiple stellar populations}\label{sec:MPs}

The $m_{\rm F322W2}$ vs.\,$m_{\rm F115W}-m_{\rm F322W2}$ CMD of
Figure\,\ref{fig:CMDs} clearly shows that the MS color broadening,
which is comparable with that expected from observational errors for
stars brighter than the MS knee, suddenly increases among M-dwarfs
fainter than the MS knee, and approaches a value of more than 0.2 mag
in the F322W2 interval between $\sim$20 and 23~mag.

\begin{figure*} 
\centering
 \includegraphics[height=12.0cm,trim={0.5cm 4.0cm 0.0cm 5.0cm},clip]{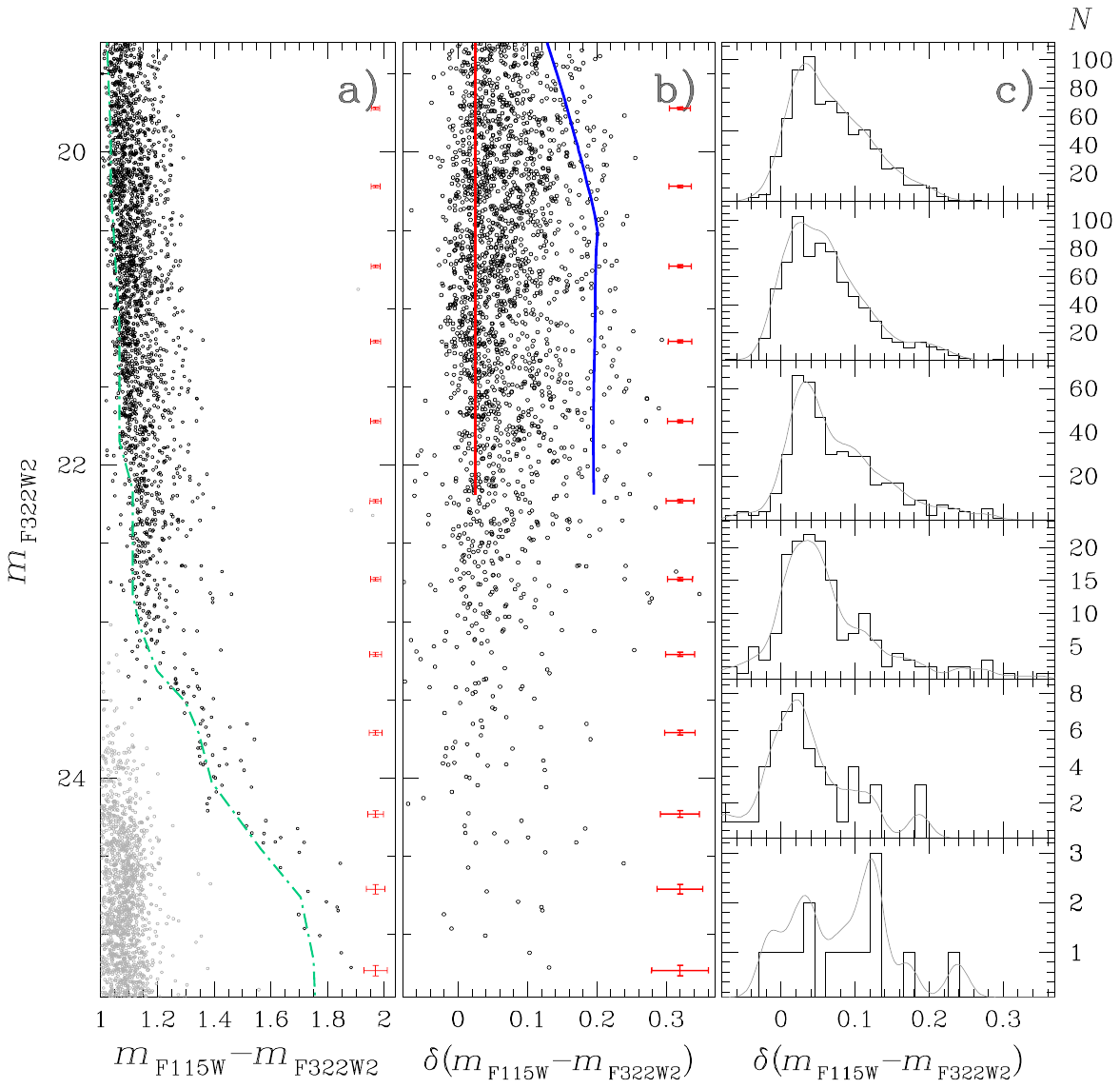}
 \caption{ Reproduction of the $m_{\rm F322W2}$ vs.\,$m_{\rm F115W}-m_{\rm F322W2}$ CMD of Figure\,\ref{fig:CMDs} zoomed around the bottom of the MS of 47\,Tucanae (panel a). 
 The aqua line is the blue boundary of the MS and is used to derive the $m_{\rm F322W2}$ vs.\,$\delta$($m_{\rm F115W}-m_{\rm F322W2}$) verticalized diagram shown in panel b. 
 The red and blue lines are the best-fit isochrones for MS stars with [O/Fe]=0.4 and [O/Fe]=$-0.1$ dex, respectively \citep{dotter2008a, milone2023a}. 
 Panels c illustrate the $\delta$($m_{\rm F115W}-m_{\rm F322W2}$)  histogram 
 distributions for stars in six F322W2 magnitude intervals and the corresponding 
 kernel-density distributions.}
 \label{fig:hist} 
\end{figure*}

Figure\,\ref{fig:hist} provides additional information on the color distribution of MS stars.
We derived by eye the aqua dashed-dot line shown in panel a) to define
the blue boundary of the MS. Subsequently, we utilized this line to
generate the verticalized $m_{\rm F322W2}$ vs.\,$\delta$($m_{\rm
  F115W}-m_{\rm F322W2}$) diagram represented in panel b). To achieve
this, we subtracted the color of each star from the color of the aqua
fiducial associated with the identical F322W2 magnitude.  
 Panels c) show the histograms and the kernel-density distributions of
 $\delta$($m_{\rm F115W}-m_{\rm F322W2}$) in six intervals of F322W2
 magnitude. The latter is obtained by using a Gaussian kernel with
 dispersion, $\sigma$=0.02 mag. 

When considering a MS portion with nearly constant F322W2 magnitude,
the majority of M-dwarfs in this luminosity interval exhibits blue
colors. In particular, the histograms of Figure\,\ref{fig:hist}c
exhibit a peak around $\delta$($m_{\rm F115W}-m_{\rm F322W2}$)=0.02
mag, which corresponds to the bulk of 1P stars.  However, there is a
tail of stars that extends towards red colors. 

The color broadening is mostly due to the star-to-star oxygen variations, which are associated with multiple stellar populations. 
The main culprit is the absorption of different chemical species
containing oxygen, mainly $\rm {H_{2}O}$ (the primary absorber) and
OH. The absorption features of these molecules strongly affect the
flux in F322W2 band 
 whereas the F115W filter is poorly sensitive to oxygen variations. As
 a consequence, the 1P stars, which have higher oxygen abundances than
 the 2P, exhibit fainter F322W2 magnitudes and bluer $m_{\rm
   F115W}-m_{\rm F322W2}$ colors than stars with similar atmospheric
 parameters \citep[][]{milone2012c, milone2023a, dotter2015a,
   vandenberg2022a, ziliotto2023a}. 

To quantify the amount of oxygen that is needed to reproduce the
$m_{\rm F115W}-m_{\rm F322W2}$ MS broadening, we compared isochrones
with different oxygen abundances \citep{dotter2008a, milone2023a} with
the observed CMD. Specifically, we considered the isochrones with
[O/Fe]=0.4 and [O/Fe]=$-$0.1. To properly compare the relative colours
of 47\,Tucanae stars with those of the isochrones, we have calculated
the colour difference 
between the O-poor isochrone and the O-rich one and shifted both
verticalized isochrones by $\delta$($m_{\rm F115W}-m_{\rm
  F322W2}$)=0.02 mag, in such a way that the red isochrone is
superimposed on the bulk of 1P stars. The $\delta$($m_{\rm
  F115W}-m_{\rm F322W2}$) interval spanned by stars brighter than
$m_{\rm F322W2} \sim 22.5$ mag, which have masses larger than
$\sim$0.1~${\rm M}_{\odot}$, is consistent with an oxygen variation of
$\sim$0.5 dex, comparable with the [O/Fe] range inferred for RGB stars
from high-resolution spectroscopy \citep[e.g.][]{carretta2009a,
  dobrovolskas2014a}. We conclude that the stars in the mass interval
between $\sim$0.1 and 0.9 solar masses span a similar range of
[O/Fe]. 

Intriguingly, as noticed by \citet{milone2023a} for the stars in the
field B, the MS color broadening of stars fainter than $m_{\rm F322W2}
\sim 23$~mag appears to be much smaller than that observed for
brighter M-dwarfs in F322W2 magnitude range $\sim$20.0-22.5. 
Unfortunately, there are no available isochrones that account for
multiple stellar populations among ultracool stars. Hence, 
we can not provide a firm conclusion on whether or not this fact is
due to the lack of stellar populations with extreme chemical
composition among these very low-mass stars. Additional qualitative
information on multiple populations among stars less massive than
$\sim$0.1 solar masses is provided by the visual comparison between
the observed $m_{\rm F322W2}$ vs.\,$m_{\rm F150W2}-m_{\rm F322W2}$ CMD
shown in Figure\,\ref{fig:CMDs2} and the CMD simulated by 
\citet[][see their figure 10]{gerasimov2023a}, which accounts for the
chemical compositions of the multiple populations of 47\,Tucanae. The
simulated CMD exhibits a wide $m_{\rm F150W2}-m_{\rm F322W2}$ color
range of $\sim$0.5~mag for stars with $m_{\rm F322W2} \sim 24$~mag,
which seems in contrast with the narrow sequence of ultracool stars
with similar luminosity observed in Figure\,\ref{fig:CMDs2}. This fact
suggests a possible lack of very O-poor 2P stars among ultracool
stars.

\begin{figure*} 
\centering
 \includegraphics[width=14.0cm,trim={0.75cm 4.5cm 0cm 8.0cm},clip]{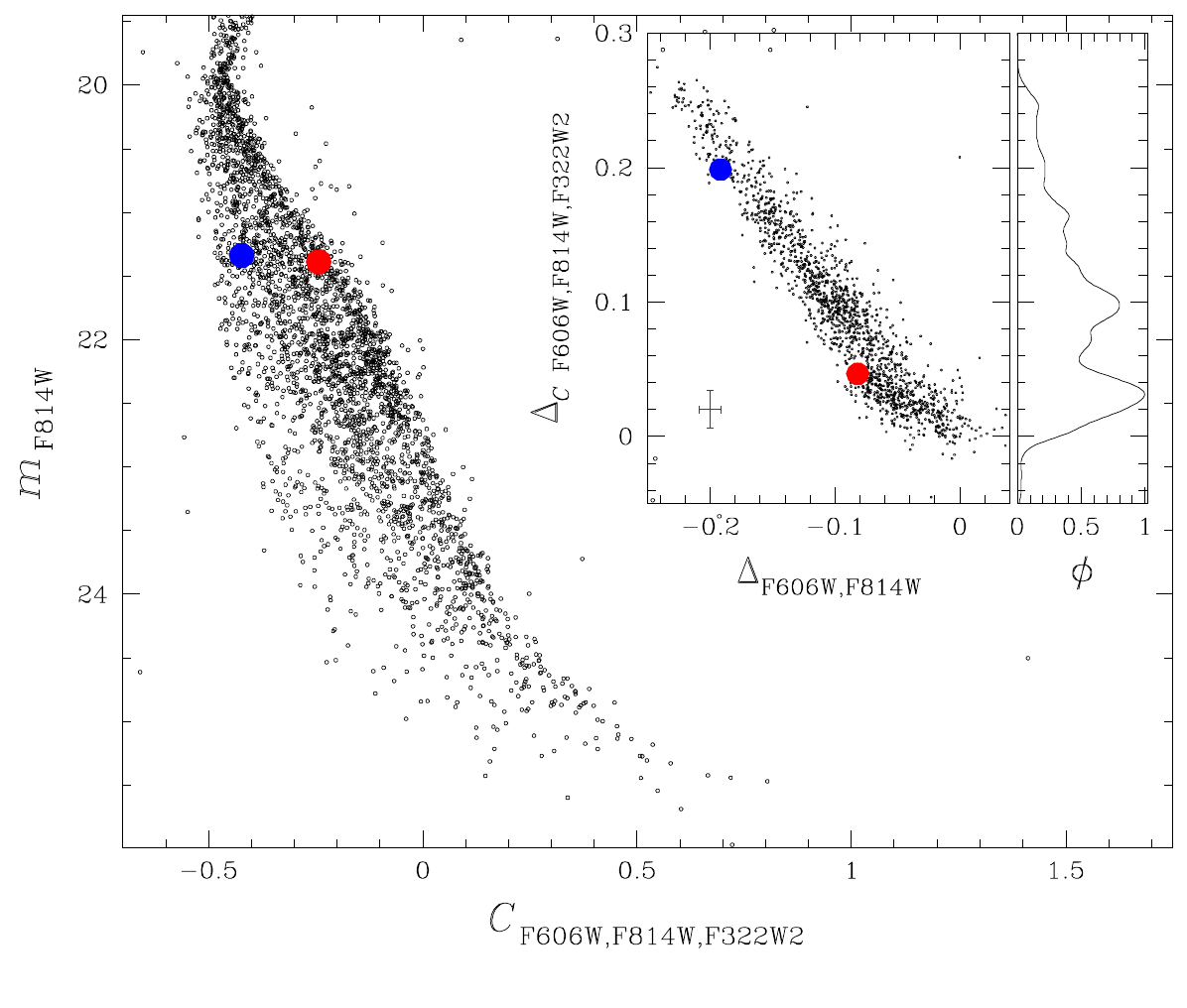}
  \includegraphics[width=8.5cm,trim={0.5cm 4.5cm 0cm 10.4cm},clip]{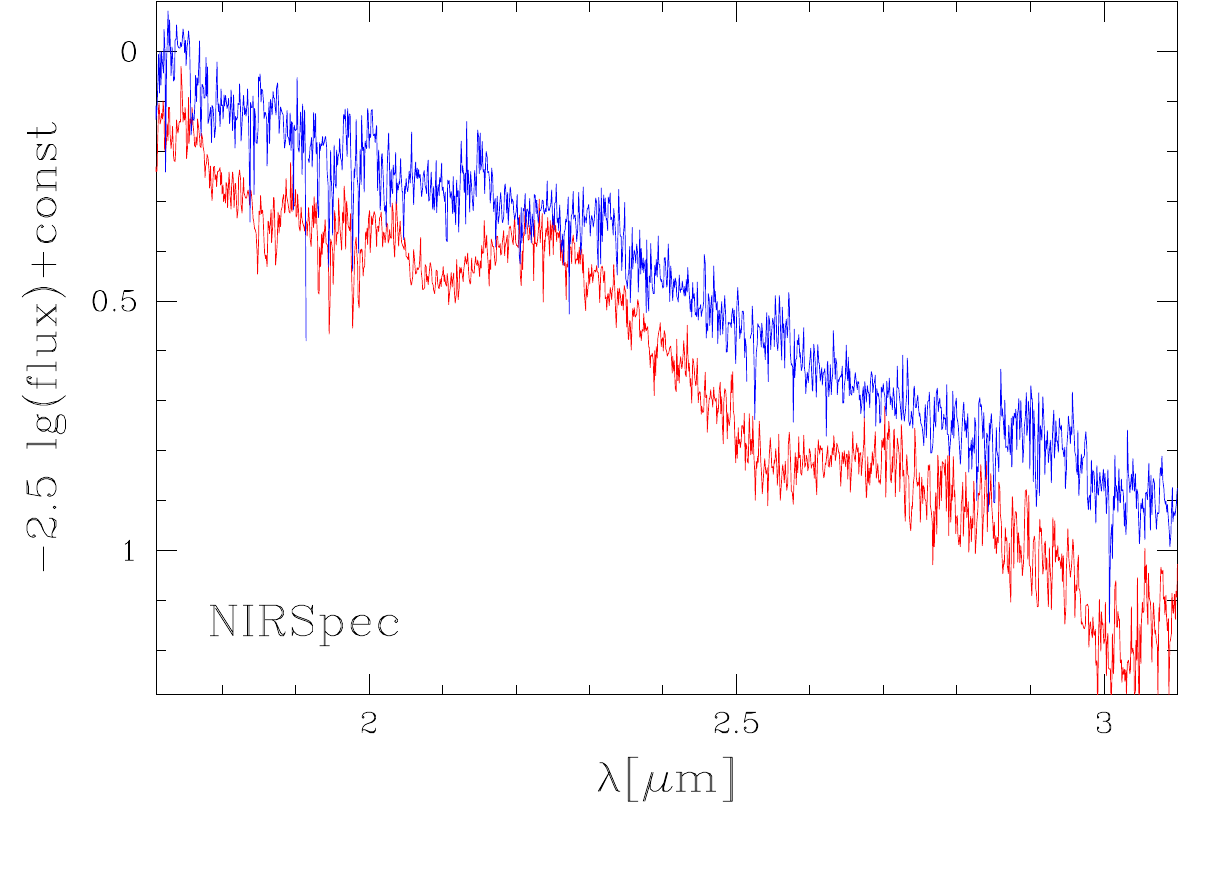}
  \includegraphics[width=8.5cm,trim={0.5cm 4.5cm 0cm 10.4cm},clip]{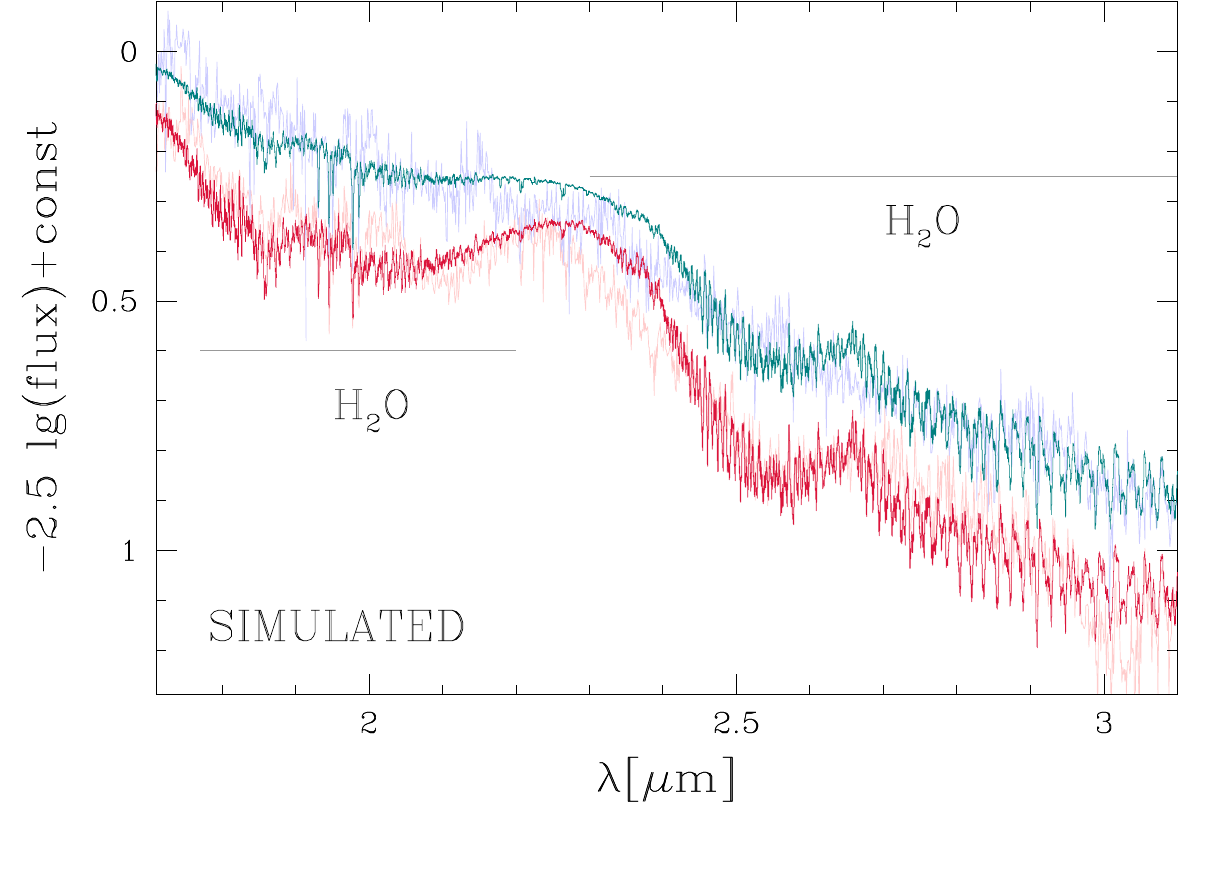}
 \caption{{\it Top.} $m_{\rm F814W}$ vs.\,$C_{\rm F606W,F814W,F322W2}$
   pseudo CMD for stars in field B. The inset shows the $\Delta_{C \rm
     F606W,F814W,F322W2}$ vs.\,$\Delta_{\rm F606W,F814W}$ ChM for MS
   stars with $21.5< m_{\rm F814W}< 24.0$ mag,
   and the corresponding $\Delta_{C \rm F606W,F814W,F322W2}$
     kernel density distribution.
{\it Bottom.} NIRSpec spectra for the two stars marked with large dots in the pseudo-CMD and in the ChM (left). Two simulated spectra with [O/Fe]=$+$0.3 (red) and [O/Fe]=$-$0.1 (blue) are superimposed on the observed spectra in the bottom-right panel. }
 \label{fig:ChMs2} 
\end{figure*}

Additional information on multiple populations in 47\,Tucanae is provided by the $m_{\rm F814W}$ vs.\,$C_{\rm F606W,F814W,F322W2}$ pseudo CMD of the stars in field B shown in the top panel of Figure\,\ref{fig:ChMs2}. In this diagram, the stars fainter than the MS-knee, identified at approximately $m_{\rm F814W} = 20.0$ mag, reveal hints of four discrete sequences. 
According to the isochrones provided by \citet{dotter2008a} and
\citet{milone2023a}, which consider the chemical composition of
multiple populations, the reddest sequence is composed of 1P stars,
while the 2P stars with particularly extreme chemical compositions
display bluer $C_{\rm F606W,F814W,F322W2}$ pseudocolors. 

To maximize the information on multiple populations, we combine the $m_{\rm F814W}$ vs.\,$C_{\rm F606W,F814W,F322W2}$ pseudo CMD and the $m_{\rm F814W}$ vs.\,$m_{\rm F606W}-m_{\rm F814W}$ CMD of M-dwarfs with $21.5< m_{\rm F814W}< 24.0$ mag.
The inset of Figure\,\ref{fig:ChMs2} shows the resulting $\Delta_{C
  \rm F606W,F814W,F322W2}$ vs.\,$\Delta_{\rm F606W,F814W}$ ChM, where
1P stars are clustered in the ChM region with $C_{\rm
  F606W,F814W,F322W2} \lesssim 0.05$ mag. We detect an extended 1P
sequence that we associate with star-to-star metallicity variations,
as found by \citet{marino2023a} based on high-resolution spectroscopy
of RGB stars \citep[see also][]{milone2017a, milone2023a,
  legnardi2022a}. 
 The 2P sequence exhibits hints of four groups of stars clustered
 around $\Delta_{\rm F606W,F814W} \sim -0.10, -0.14, -0.17$, and
 $-0.20$ mag.

\section{The chemical composition of M-dwarfs from NIRSpec spectra}\label{sec:NIRSpec}

In this Section we present NIRSpec spectra for two M dwarfs photometrically associated with different stellar populations. The location of these stars in the photometric diagrams is shown in Figure\,\ref{fig:ChMs2}.
The large dots superimposed on the pseudo-CMD and the ChM of Figure\,\ref{fig:ChMs2} mark indeed two stars with comparable F814W magnitudes. However, these stars exhibit distinct values for $\Delta_{C \rm F606W,F814W,F322W2}$ and $\Delta_{\rm F606W,F814W}$. Based on their location in Figure\,\ref{fig:ChMs2}, upper panel, the red- and blue-dot stars are labelled as 1P and 2P, respectively.

The two targets have been spectroscopically observed as part of GO-2560, and their
NIRSpec spectra are displayed in the bottom-left panel of Figure\,\ref{fig:ChMs2}. 
 The red spectrum represents the star with RA=0$^{h}$ 22$^{m}$ 22.14$^{s}$ and DEC=$-72^{d}$ 04$^{m}$ 10.5$^{s}$ while the blue spectrum correspond to the star with RA=0$^{h}$ 22$^{m}$ 12.36$^{s}$ and DEC=$-72^{d}$ 04$^{m}$ 30.8$^{s}$.
 In presenting the spectra, we applied the same shift to both spectra, plotting them such that the average value of $-$2.5 log$_{10}$(flux) in the region with $\lambda<1.75$~$\mu$m for the blue spectrum equals zero.

Clearly, the spectrum of the 1P star exhibits much stronger molecular bands than that of the 2P star. 
In principle, such large difference in the  spectra could be either due to different atmospheric parameters or to different chemical composition. The location of the two stars at similar $m_{\rm F814W}$ magnitudes suggests that they have similar atmospheric parameters. Indeed, by fitting the $m_{\rm F814W}$
vs.\,$m_{\rm F606W}-m_{\rm F814W}$ CMD with the isochrones by \citet{dotter2008a} and \citet{milone2023a} we obtain \teff/\logg=3700/5.0 and 3600/5.0 for the blue and the red star, respectively. Such a small difference in the atmospheric parameters alone cannot account for the differences  of the observed spectra.

 As outlined in the spectra of Figure\,\ref{fig:ChMs2}, the main molecular feature appearing in the observed spectral range are due to rotational-vibrational ${\rm {H_{2}O}}$ band heads at 1.9 and 2.7~$\mu$m, and some contribution from the CO second overtone bands are rather masked by the ${\rm {H_{2}O}}$ bands.
The water vapour molecular bands in the near-IR are good indicators of oxygen abundance. 

While for a full detailed spectroscopic analysis of all the observed NIRSpec spectra we refer the reader to an upcoming paper, here we illustrate two synthetic spectra constructed with the atmospheric parameters assumed for our two M dwarfs to infer a first estimate of the O range in the low mass regime of 47~Tucanae.

The bottom-right panel of Figure\,\ref{fig:ChMs2} shows the comparison of the observed spectra with simulated spectra with different light-element abundances. 
The simulated spectra are derived as in our previous works \citep{milone2023a, ziliotto2023a}. In a nutshell, we calculated the model atmospheres using the ATLAS 12 computer program, employing the opacity-sampling technique and assuming local thermodynamic equilibrium \citep{kurucz1970a, sbordone2004a}. We incorporated molecular line lists for all diatomic molecules listed on Kurucz's website, in addition to including H$_2$O molecules from \citet{partridge1997a}. We used the SYNTHE computer program \citep{kurucz1981a, castelli2005a, kurucz2005a, sbordone2007a}
 to compute the spectra in the region between $1.7$ and 3.2 $\mu$m covered by the available NIRSpec spectra.
 Specifically, the crimson spectrum has oxygen content [O/Fe]=0.3 dex, whereas the teal one has [O/Fe]=$-$0.1~dex.
 Noticeably, the blue spectrum has higher effective temperature than the red one by 70~K. This effective temperature difference is consistent with the fact that 2P stars of 47\,Tucanae are helium enhanced with respect to the 1P \citep[e.g.\,][]{lagioia2018a, milone2018a}.

 We conclude that the comparison of the spectra of the two M dwarfs belonging to different stellar populations indicates that M dwarfs exhibit  a range in the O abundances very similar to that reported in the literature on higher mass stars along the RGB. The result presented here is the first detection of the O range directly performed on stellar spectra of M dwarfs, and strongly corroborates our previous findings about the similarity of the multiple stellar populations in stars with different masses (from the RGB to M dwarfs), already suggested by photometric diagrams.

\section{Population ratios}\label{sec:pratio}

    To derive the fraction of stars in the main stellar populations of 47\,Tucanae we started using 
    the $\Delta_{F110W,F160W,F115W,F322W2}$ vs.\,$\Delta_{F606W,F110W}$
     ChM of stars in field C, which is derived from the $m_{\rm
       F322W2}$ vs.\,$m_{\rm F110W}-m_{\rm F160W}+m_{\rm F115W}-m_{\rm
       F322W2}$ pseudo-CMD plotted in Figure\,\ref{fig:PMs} and the
     $m_{\rm F322W2}$ vs.\,$m_{\rm F606W}-m_{\rm F110W}$ CMD.  
     
\begin{centering} 
\begin{figure*} 
 \includegraphics[height=10.0cm,trim={0.5cm 4.5cm 7cm 6.0cm},clip]{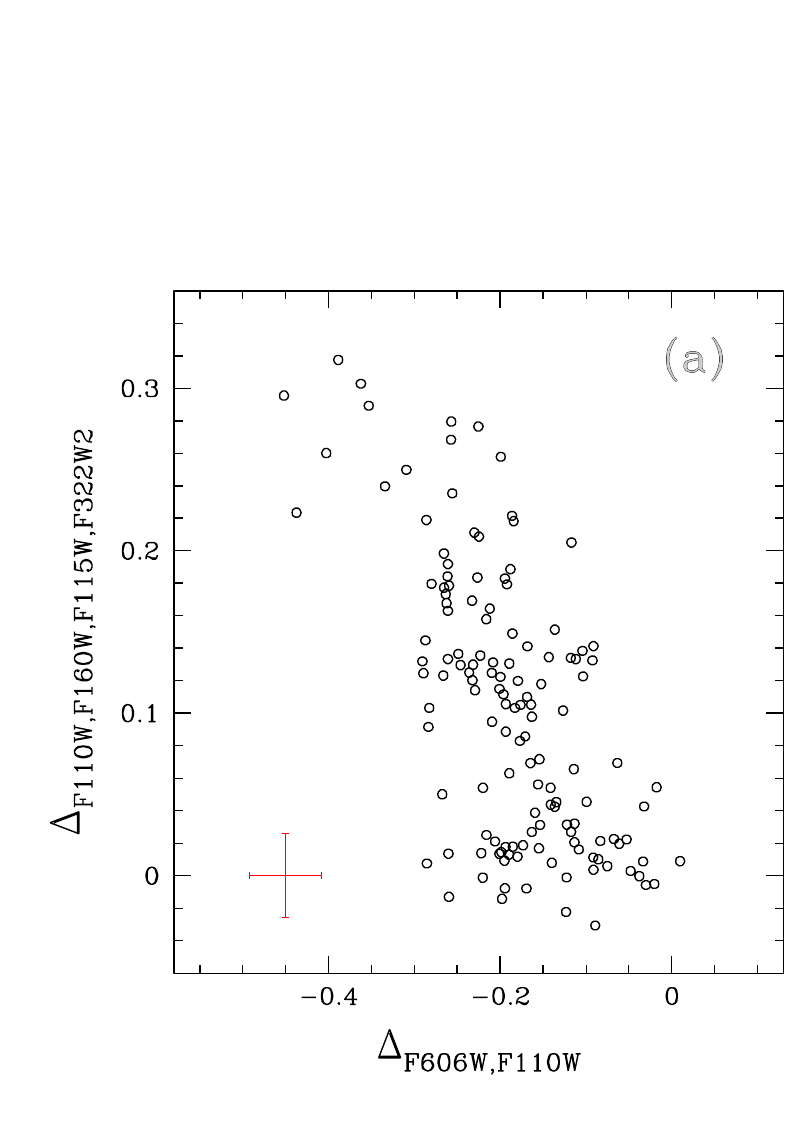}
  \includegraphics[height=10.0cm,trim={0.5cm 4.5cm 2cm 6.0cm},clip]{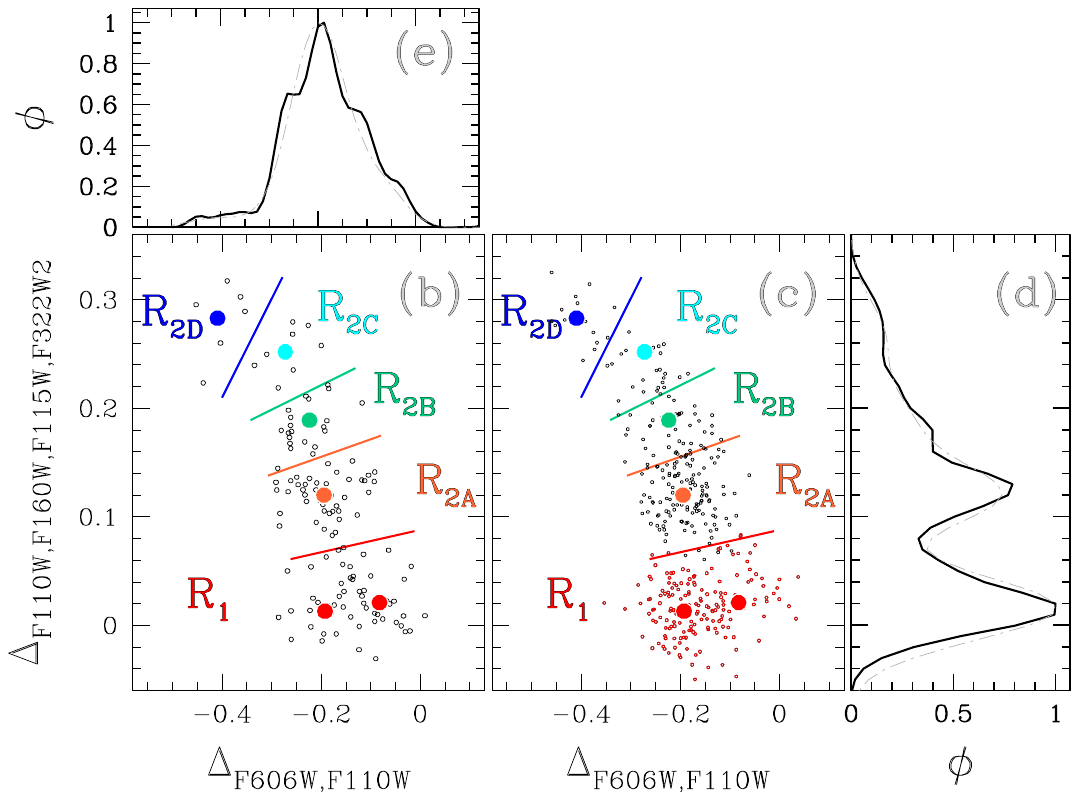}
 \caption{ $\Delta_{F110W,F160W,F115W,F322W2}$ vs.\,$\Delta_{F606W,F110W}$ ChM of M-dwarfs in field C (panel a). Panels b--e illustrate the procedure to estimate the fractions of 1P and 2P stars. The observed ChM is further reproduced in panel (b), while panel (c) shows the simulated ChM where simulated 1P stars are coloured in red. The colored solid lines delimit the four regions used to infer the fraction of stars in each population, while the colored dots mark the average positions in the ChM of the simulated stellar populations. Panels d and e show the $\Delta_{F110W,F160W,F115W,F322W2}$ and $\Delta_{F606W,F110W}$ kernel distributions for the observed (black line) and the simulated (grey line) stars.
  }
 \label{fig:ChMs} 
\end{figure*} 
\end{centering}

     The results are plotted in the left panel of Figure\,\ref{fig:ChMs}. 
    The 1P stars are clustered around the origin of the ChM, while the 2P defines a sequence ranging from ($\Delta_{F606W,F110W}$,$\Delta_{F110W,F160W,F115W,F322W2}$) ($\sim -$0.2, 0.1) towards the top left corner of the ChM.

To assess the proportion of stars within each population, we expanded upon the approach outlined by \citet{zennaro2019a} to derive the fraction of stars in the four stellar populations of the GC NGC\,2419 and apply it to the ChM of 47\,Tucanae \citep[see also][]{milone2012a, milone2020a, nardiello2018a}. 
 In a nutshell, we estimated the mean values of $\Delta_{\rm F606W,F110W}$ and $\Delta_{F110W,F160W,F115W,F322W2}$ for the stars within each population (depicted as colored dots in Figure\,\ref{fig:ChMs}). In the case of 1P stars, we used two values for the center to account for sub-populations with different metallicity \citep{legnardi2022a, marino2023a}. 
  These values served to define five distinct regions denoted as R$_{1}$, R$_{\rm 2A}$, R$_{\rm 2B}$, R$_{\rm 2C}$, and R$_{\rm 2D}$ and delimited by the colored solid lines.
Owing to photometric errors, each region may encompass stars from all populations. As an example, the observed count of stars within region R1 is:
\begin{equation}
    N_{\rm R1}=N_{\rm 1P}f^{\rm R1}_{\rm 1P} + N_{\rm 2PA}f^{\rm R1}_{\rm 2PA} + N_{\rm 2PB}f^{\rm R1}_{\rm 2PB} + N_{\rm 2PC}f^{\rm R1}_{\rm 2PC}+ N_{\rm 2PD}f^{\rm R1}_{\rm 2PD}
\end{equation}

where N$_{\rm 1P}$, N$_{\rm 2PA}$, N$_{\rm 2PB}$, N$_{\rm 2PC}$, and N$_{\rm 2PD}$, are the numbers of 1P, 2P$_{\rm A}$, 2P$_{\rm B}$, 2P$_{\rm C}$, and 2P$_{\rm D}$ stars in each population in our sample.

The number of stars within the regions R$_{\rm 2A}$, R$_{\rm 2B}$, R$_{\rm 2C}$, and R$_{\rm 2D}$ are linked to the fraction of stars of each population through four comparable equations.
The values of N$_{\rm R1}$, $N_{\rm R2A}$, $N_{\rm R2B}$, $N_{\rm R2C}$, and $N_{\rm R2D}$ used in these equations are derived by counting the stars within the corresponding regions. 

The fractions of 1P, 2P$_{\rm A}$, 2P$_{\rm B}$, 2P$_{\rm C}$, and 2P$_{\rm D}$ stars within the region R1, $f^{\rm R1}_{\rm 1P}$, $f^{\rm R1}_{\rm 2PA}$, $f^{\rm R1}_{\rm 2PB}$, $f^{\rm R1}_{\rm 2PC}$, and  $f^{\rm R1}_{\rm 2PD}$ are derived from simulated ChMs composed of ASs, and we did the same for inferring the fractions of stars of each populations in the regions of the ChM R$_{\rm 2A}$, R$_{\rm 2B}$, R$_{\rm 2C}$, and R$_{\rm 2D}$.
 To do this, we simulated 50,000 ASs for each population, disposed along the centers of each population in the ChM. 

The numbers of stars in the five populations of 47\,Tucanae, $N_{\rm 1P}, N_{\rm 2PA}, N_{\rm 2PB}, N_{\rm 2PC}$, $N_{\rm 2PD}$, are calculated by solving for these five equations. The results are provided in Table\,\ref{tab:pratio}, where we also provide the fractions of stars in the five stellar populations that we obtain by extending the same procedure to the ChM shown in Figure\,\ref{fig:ChMs2} for stars in field A. In particular, we find that 1P stars comprise the 38.1$\pm$1.2\% and the 46.3$\pm$4.2\% of the total number of ChM stars of field A and field C, respectively.

\begin{table*}
  \caption{Fractions of stars in the five stellar populations of 47\,Tucanae for the fields A and C.}
\label{tab:pratio}
\centering
\begin{tabular}{l c c c c c }
\hline \hline
Field & 1P & 2P$_{\rm A}$ & 2P$_{\rm B}$  & 2P$_{\rm C}$ & 2P$_{\rm D}$   \\
 A    & 0.381$\pm$0.012 &  0.352$\pm$0.012  & 0.148$\pm$0.009 & 0.065$\pm$0.006 & 0.054$\pm$0.006 \\
 C    & 0.463$\pm$0.042 &  0.329$\pm$0.035  & 0.122$\pm$0.028 & 0.049$\pm$0.021 & 0.036$\pm$0.014   \\

     \hline\hline
\end{tabular}
 \end{table*}

\begin{centering} 
\begin{figure} 
 \includegraphics[width=8.5cm,trim={0.5cm 5cm 0cm 12.0cm},clip]{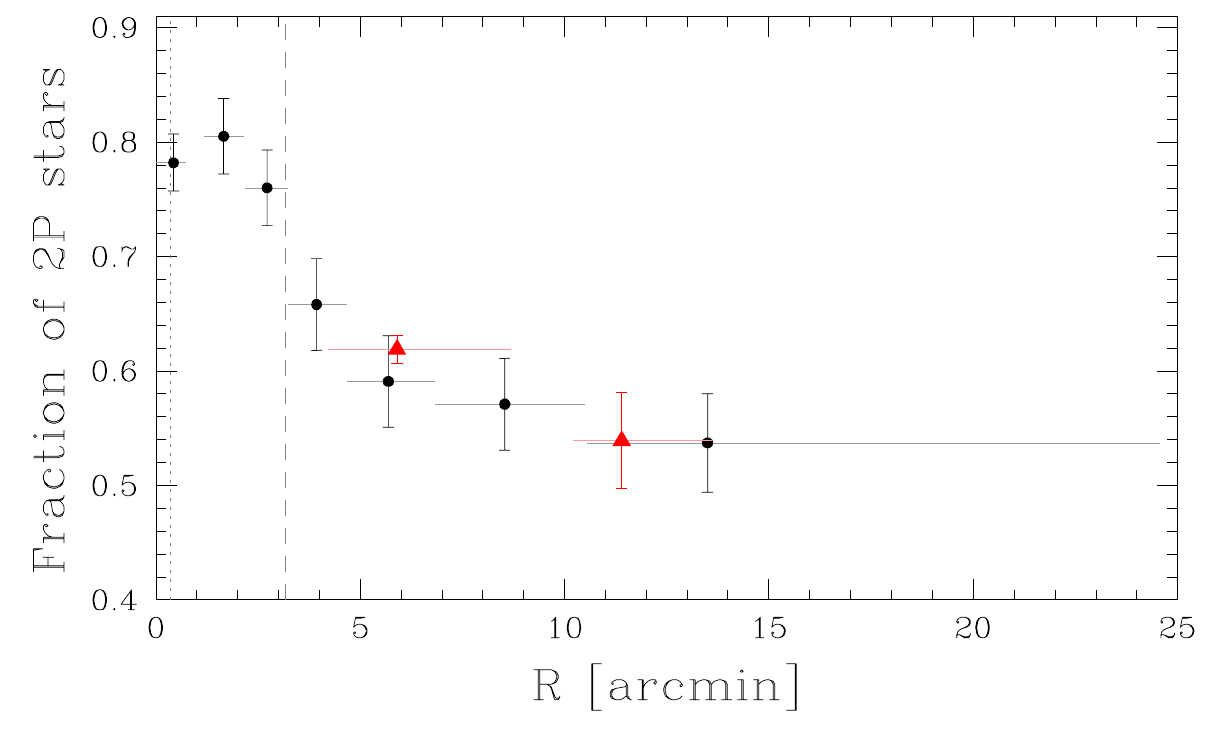}
 \caption{Fraction of 2P stars as a function of the radial distance from the cluster centers. Black dots are derived by \citet{dondoglio2021a} by using HB stars, whereas the results obtained in this paper from M-dwarfs are represented with red triangles. The horizontal lines mark the radial interval associated with each point. The dotted and dashed vertical lines mark the core and the half-light radii from the 2010 version of the \citet{harris1996a} catalog.}
 \label{fig:RD} 
\end{figure} 
\end{centering}

The fraction of 2P stars in 47\,Tucanae inferred from either RGB or HB stars significantly changes as a function of the radial distance. It ranges from about 80\% near the cluster centers to less than 60\% for distances larger than $\sim$5~arcmin \citep{milone2012b, milone2017a, cordero2014a, dondoglio2021a, lee2022a}.
 Figure\,\ref{fig:RD} compares the values of the fractions of 2P stars
 derived by \citet{dondoglio2021a} from HB stars with different radial
 distances (black points) and those obtained in this paper from
 M-dwarfs (red triangles). We conclude that there is no significant
 difference between the fractions of 2P stars derived from stars with
 different masses. 
In the context of the multiple generations scenario, where the 2P
stars form in the dense cluster center, this would imply that the IMF
does not depend on the density of the environment.

\section{Summary and conclusions}\label{sec:summary}

We have presented our $JWST$ project on multiple stellar populations in the GC 47~Tucanae. 
Our main dataset comprised $\sim$13h observing time with NIRSpec$+$NIRCam parallel observations, allowing us to infer crucial constraints on the multiple-population phenomenon both from spectroscopic and photometric analysis.
This is the first project specifically devoted to the analysis of multiple stellar populations in GCs conducted with the $JWST$.
However, in addition to studies on multiple populations, our dataset demonstrated to have the potential to investigate the general properties of the poorly-explored stellar populations, such as the M-dwarfs and brown-dwarfs.
In this work, we present early results which allowed us to study the
low mass stars, in the domain of M dwarfs and beyond, down to the
hydrogen-burning limit and the brown-dwarf sequence. 

This overview sheds light on multiple populations among low-mass stars of 47\,Tucanae. The main results can be summarized as follows.

\begin{itemize}

\item{The photometric diagrams constructed with the F322W2 filter of NIRCam, including the $m_{\rm F322W2}$ vs.\,$m_{\rm F115W}-m_{\rm F322W2}$ CMD and the $m_{\rm F322W2}$ vs.\,$C_{\rm F606W,F814W,F322W2}$ pseudo-CMD,  reveal that, below the knee, MS stars with masses larger than $\sim 0.1$ M$_{\odot}$ and similar F322W2 luminosities span a wide color, or pseudo color, range. This result corroborates the evidence of multiple stellar populations among M-dwarfs \citep[see e.g.,][for previous photometric studies of multiple populations among very low mass stars in 47 Tucanae]{dondoglio2022a, milone2022a, milone2023a, cadelano2023}.} 

\item{The comparison between isochrones with different chemical compositions and the observed CMD reveals that the MS color broadening observed among M-dwarfs more-massive than $\sim$0.1 solar masses is consistent with stellar populations with different oxygen abundances. The [O/Fe] range of $\sim$0.5 dex that is needed to reproduce the observations is comparable with that observed among RGB stars by means of high-resolution spectroscopy. We notice that the $m_{\rm F115W}-m_{F322W2}$} color broadening of very-faint stars ($m_{\rm F322W2}> 23$ appears to be narrower than that observed among the M-dwarfs with brighter luminosity, similar with what was observed by \citet{milone2023a} for stars in field B. An appropriate comparison with isochrones that accounts for the chemical compositions of multiple populations is needed to associate this fact with the possible lack of stellar populations with extreme chemical compositions among ultracool stars.

\item{We present the NIRSpec spectra for two stars with similar F814W magnitude that occupy extreme positions in the ChM and are associated with the populations 1P and 2P$_{\rm C}$. For a fixed wavelength, the 1P-star spectrum is more-absorbed than the spectrum of the 2P$_{\rm C}$ star. The comparison with synthetic spectra with different chemical compositions reveals that the two stars have different oxygen abundances, with the 2P$_{\rm C}$ M-dwarf being depleted by [O/Fe]=0.4 dex with respect to the 1P star. 
Hence, the flux difference can be attributed to molecules consisting of oxygen (predominantly H$_2$O), which exhibit strong absorption in the spectra of 1P stars, known for their oxygen-rich composition.
This outcome represents the first spectroscopic determination of the chemical composition of M-dwarfs within a GC. It validates the earlier prediction, derived from synthetic spectra, that the observed color variation among M-dwarfs in GCs is attributable to multiple stellar populations characterized by varying oxygen abundances \citep{milone2012c}. }

\item{The 2P M dwarfs consists in $\sim$62\% and 54\% of the total
    number of M dwarfs, respectively at $\sim 5$ and $\sim 11$~arcmin from the cluster center. These values are similar to the fraction of 2P stars measured among HB and RGB stars with similar radial distance \citep{dondoglio2021a, milone2012b}, thus indicating that the fractions of 2P stars do not depend on stellar mass, at least for stars more massive than $\sim$0.1\,M$_{\odot}$.
This result, together with the evidence that M-dwarfs and giant stars span a similar range of [O/Fe], 
provides a serious challenge to the scenarios for the formation of multiple populations that are based on accretion.
}
\end{itemize}

The deep NIRCam photometry collected as part of GO-2560 has allowed us to detect objects in the field of 47\,Tucanae down to $m_{\rm F322W2}=27.0$. We have thus explored the faintest MS  and brown dwarf regions, which are poorly investigated in GCs. The corresponding results can be summarised as follows:
    
\begin{itemize}
\item{Based on both the $m_{\rm F322W2}$ vs.\,$m_{\rm F115W}-m_{\rm F322W2}$ and the $m_{\rm F322W2}$ vs.\,$m_{\rm F150W2}-m_{\rm
      F322W2}$ CMDs, we detected a main discontinuity along the
      sequence of very low mass stars (masses smaller than
      $\sim$0.1\,M$_{\odot}$) at $m_{\rm F322W2} \sim 24.2$~mag. We
      also notice clear changes of the MS slope around $m_{\rm
        F322W2}=23.3$ and $24.5$~mag. 
}

\item{The F322W2 luminosity function of MS stars exhibits a drop in
    the number of stars around $m_{\rm F322W2}=25.3$ mag, and the
    number of stars per magnitude interval rises up at fainter
    luminosities. We tentatively associate this gap with the hydrogen-burning limit. Hence, the deep CMD obtained from GO-2560 photometry unveils, for the first time in a GC, the brown dwarfs cooling sequence.}

\end{itemize}

We can conclude by emphasising that the initial observations of GCs carried out with the $JWST$ have showcased the telescope's capability to effectively separate multiple populations among M dwarfs and explore the properties of very low mass stars. The results mark the opening of new horizons in the exploration of the lower mass regime. The heightened sensitivity of
$JWST$ to the infrared domain will be indeed instrumental in enabling an
efficient and systematic exploration of cool, low-mass stars across a substantial sample of GCs. 
A current limitation is represented by the insufficient availability of theoretical isochrones and mass-luminosity relations down to the faint limits now reached with $JWST$.

\section*{Acknowledgments} 
\small
We thank the anonymous referee for his/her suggestions, that improved
the manuscript.
This work has received funding from 
"PRIN 2022 2022MMEB9W - {\it Understanding the formation of globular
  clusters with their multiple stellar generations}", from 
INAF Research GTO-Grant Normal RSN2-1.05.12.05.10 -  (ref. Anna
F. Marino) of the "Bando INAF per il Finanziamento della Ricerca
Fondamentale 2022", from the European Union’s Horizon 2020
research and innovation programme under the Marie
Skłodowska-Curie Grant Agreement No. 101034319 and from the
European Union – NextGenerationEU. 
YC acknowledges support from the grant RYC2021-032718-I, financed by
MCIN/AEI/10.13039/501100011033 and the European Union 
NextGenerationEU/PRTR. This work has been partially supported by the
Spanish MINECO grant PID2020-117252GB-I00 and by the ABAUR/Generalitat 
de Catalunya grant SGR-386/2021.
SJ acknowledges support from the NRF of Korea (2022R1A2C3002992, 2022R1A6A1A03053472).
\section*{Data availability}
The data underlying this article will be shared on reasonable request to the corresponding author. \\

\bibliography{47tuc_accepted.bbl}

\end{document}